\documentclass[12pt,lettersize,journal,onecolumn]{IEEEtran}
\usepackage[T1]{fontenc}% optional T1 font encoding
\usepackage{bm}
\usepackage{color}
\usepackage{amsmath}
\usepackage{url}
\usepackage{array}
\usepackage{booktabs}
\usepackage{fancyhdr}
\usepackage{latexsym}
\usepackage{diagbox}
\usepackage{epsfig}
\usepackage{epstopdf}
\usepackage{amssymb}
\usepackage{amsfonts}
\usepackage{amsxtra}
\usepackage{xspace}
\usepackage{makeidx}
\usepackage{graphics}
\usepackage{theorem}
\usepackage{subfigure}
\usepackage{multirow}
\usepackage{verbatim}
\usepackage{algorithmic,algorithm}

\usepackage[noadjust]{cite}
\usepackage{caption}
\usepackage{subcaption}
\usepackage{stfloats} % 增强浮动对象控制
\usepackage{setspace}
\usepackage[caption=false,font=normalsize,labelfont=sf,textfont=sf]{subfig}
\newtheorem{lemma}{\textbf{Lemma}}

\begin{document}

\title{Heterogeneous-IRS-Assisted MIMO Systems: Channel Estimation and Beamforming}

\author{Weibiao~Zhao,~\IEEEmembership{Graduate~Student~Member,~IEEE,}
        Qiucen~Wu, Yuanqi~Tang,
        and~Yu~Zhu,~\IEEEmembership{Senior~Member,~IEEE}
        
\thanks{Weibiao Zhao, Qiucen Wu, Yuanqi Tang, and Yu Zhu are with the Key Laboratory for Information Science of Electromagnetic Waves (MoE), School of Information Science and Technology, Fudan University, Shanghai 200433, China (e-mail: zhaowb22@m.fudan.edu.cn; qcwu21@m.fudan.edu.cn; yqtang22@m.fudan.edu.cn; zhuyu@fudan.edu.cn).}
\thanks{This work was supported by the Natural Science Foundation of Shanghai under Grant No. 23ZR1407300, and the National Natural Science Foundation of China under Grant No. 62471145. \textit{(Corresponding author: Yu Zhu.)}}
}

\maketitle

\begin{abstract}
Intelligent reflecting surface (IRS) has gained great attention for its ability to create favorable propagation environments. However, the power consumption of conventional IRSs cannot be ignored due to the large number of reflecting elements and control circuits. To balance performance and power consumption, we previously proposed a heterogeneous-IRS (HE-IRS), a green IRS structure integrating dynamically tunable elements (DTEs) and statically tunable elements (STEs). Compared to conventional IRSs with only DTEs, the unique DTE-STE integrated structure introduces new challenges in both channel estimation and beamforming. In this paper, we investigate the channel estimation and beamforming problems in HE-IRS-assisted multi-user multiple-input multiple-output systems. Unlike the overall cascaded channel estimated in conventional IRSs, we show that the HE-IRS channel to be estimated is decomposed into a DTE-based cascaded channel and an STE-based equivalent channel. Leveraging it along with the inherent sparsity of DTE- and STE-based channels and manifold optimization, we propose an efficient channel estimation scheme. To address the rank mismatch problem in the imperfect channel sparsity information, a robust rank selection rule is developed. For beamforming, we propose an offline algorithm to optimize the STE phase shifts for wide beam coverage, and an online algorithm to optimize the BS precoder and the DTE phase shifts using the estimated HE-IRS channel. Simulation results show that the HE-IRS requires less pilot overhead than conventional IRSs with the same number of elements. With the proposed channel estimation and beamforming schemes, the green HE-IRS achieves competitive sum rate performance with significantly reduced power consumption.
\end{abstract}


\section{Introduction}\label{sec:Introduction}
Intelligent reflecting surface (IRS) has been recently recognized as a key enabling technology in the development of 6G communication \cite{2021_Pan_RIS_6G}. Typically, an IRS is a two-dimensional passive electromagnetic metasurface consisting of numerous reconfigurable reflecting elements that can be dynamically controlled. By smartly tuning the electromagnetic responses of these elements, IRSs can manipulate incident signals to reconfigure the propagation channels between the base station (BS) and user equipment (UE), thereby enhancing wireless system coverage and throughput \cite{2019_Wu_RIS}. Moreover, the passive nature of IRS elements leads to lower power consumption and hardware costs compared to traditional multi-input multi-output (MIMO) systems with active devices \cite{2018_Hu_RIS_beyondMIMO}.

Despite these advantages, the large number of reflecting elements in IRSs can still result in significant power consumption and increased hardware costs. For instance, in commonly used IRS designs based on positive-intrinsic-negative (PIN) diodes, recent experimental results show that 3,600 1-bit reflecting elements can consume up to 45 W of power \cite{2024_Jin_Power}. This level of power consumption is no longer negligible when compared to the transmit power of the BS in the downlink or the UE in the uplink. Additionally, these reflecting elements also imposes high hardware demands on the IRS controller, including the configurations of the field programmable gate array (FPGA) control board and the corresponding driving circuits \cite{2023_Rana_RIS_hardware}.
% , which increases the hardware complexity due to the need for data processing and circuit configuration \cite{2023_Rana_RIS_hardware}.

The issue of IRS power consumption has received growing attention. In \cite{2019_Huang_RIS_EE}, IRS power consumption was modeled as a function of the number of reflecting elements and the precision of their phase shifts, and was then incorporated into energy efficiency (EE) optimization for IRS-assisted systems. Building on this model, the authors of \cite{2021_You_RIS_resource_efficiency} proposed a resource efficiency optimization framework for balancing EE and spectral efficiency (SE). Distributed IRS systems were explored in \cite{2022_Yang_distributed_RIS_EE} to maximize EE by switching IRSs on or off.
% In this work, it is assumed that IRSs in the on-state serve UEs but consume constant power, while those in the off-state consume no power and do not reflect signals.
Besides, it was further revealed in \cite{2024_Jin_Power} that PIN diodes consume significant power when switched on, while requiring no power when switched off. 
% the power consumption differences between the on and off states of IRS elements were further analyzed in \cite{2024_Jin_Power}, revealing that PIN diodes consume significant power when switched on, whereas their power consumption is negligible when switched off.
Based on this finding, the relationship between IRS phase shifts and their dynamic power consumption was modeled as a phase shift-dependent power consumption (PS-DPC) model in \cite{2023_Li_RIS_EE}\cite{2025_Wu_RIS_Rate}. Using the PS-DPC model, the authors of \cite{2023_Li_RIS_EE} investigated the EE maximization problem in IRS-assisted systems. Furthermore, the authors of \cite{2025_Wu_RIS_Rate} studied the rate maximization problem by jointly optimizing beamforming and power allocation between the BS and IRS, and proposed a flexible power allocation strategy to balance the BS and IRS beamforming quality.

\begin{figure}
	\centering
    \captionsetup{font=small}
	\includegraphics[width=0.6\columnwidth]{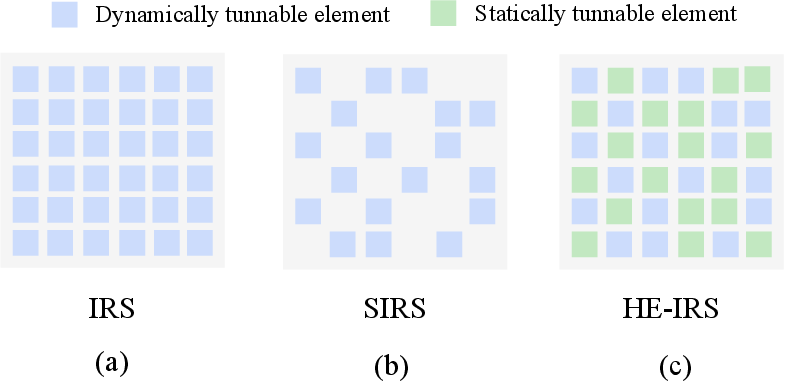}
	\caption{Examples of element compositions for different IRS structures. (a) Conventional IRS. (b) SIRS. (c) HE-IRS.}	\label{fig:structure}
\end{figure}

Most existing works focused on balancing performance and power consumption in IRS-assisted systems through optimization algorithms. However, developing new green IRS structures to achieve this balance is also important. Compared to the conventional IRS with fully populated elements in Fig. 1(a), a sparse IRS (SIRS) structure was proposed in \cite{2022_Aurangzeb_SIRS}, where reflecting elements are sparsely distributed across the surface, as shown in Fig. 1(b). While this structure reduces total power consumption as sparsity increases, it also results in performance degradation with less elements employed. To address this limitation, in \cite{2024_Zhao_HEIRS}, we proposed a green heterogeneous-IRS (HE-IRS) structure in Fig. 1(c). The HE-IRS integrates two types of reflecting elements, dynamically tunable elements (DTEs), similar to those used in the conventional IRS, and statically tunable elements (STEs). The static phase-tuning capability of STEs  enables their implementation using pre-designed structures that consume no power.
% allow them to be implemented through pre-designed structures with no power, eliminating the need for active components or control circuits required by DTEs.
Thus, the HE-IRS establishes a new trade-off between performance and power consumption \cite{2024_Zhao_HEIRS}.
% It has been demonstrated that in single-user scenarios, an HE-IRS with half of its elements as DTEs and the other half as STEs can achieve a SE comparable to that of a conventional IRS with all DTEs, but with much lower power consumption.
However, the study in \cite{2024_Zhao_HEIRS} assumed perfect channel state information (CSI) for beamforming, which is too ideal in practical systems. Moreover, only single-user scenarios were considered.
% However, when this unique DTE-STE integrated structure with both dynamic and static phase-shifting capabilities meets more complex multi-user scenarios, investigating the HE-IRS beamforming characteristics and designing tailored beamforming algorithms remain critical tasks for HE-IRS-assisted systems.

On the other hand, accurate CSI is crucial for the design of IRS beamforming. Due to the lack of baseband processing capabilities in IRS elements, channel estimation in IRS-assisted systems remains a challenging task. Several methods have been proposed in recent works. In \cite{2019_Mishra_IRS_channel_estimation}, the cascaded channel for each IRS element was sequentially estimated by adopting the least squares (LS) method. Using the LS estimation, the authors of \cite{2022_Jensen_RIS_LS} investigated the overall cascaded channel estimation when all IRS elements are activated simultaneously. By treating received pilots as tensors, a parallel factor decomposition algorithm was proposed in \cite{2021_Araújo_IRS_channel_estimation} for IRS cascaded channel estimation. Given the large matrix dimensions of the cascaded channel, 
% determined by the product of the number of the DTEs at the IRS, the number of the antennas at the BS, and those at the UE,
various compressed sensing (CS) methods exploiting channel sparsity have been proposed to reduce pilot overhead, including approximate message passing (AMP) \cite{2022_Wei_IRS_channel_estimation_AMP}\cite{2021_Mirza_cascaded_channel_CS}, orthogonal matching pursuit (OMP) \cite{2022_Lin_channel_estimation}\cite{2020_Wang_IRS_channel_estimation_OMP}, and atomic norm minimization (ANM) \cite{2022_He_cascaded_channel_ANM}\cite{2022_Liu_ANM_channel_estimation}, etc. For instance, the authors of \cite{2021_Mirza_cascaded_channel_CS} proposed an extended AMP algorithm to estimate ill-conditioned channel matrices in IRS-assisted systems and improve estimation accuracy by eliminating permutation ambiguities. 
% In \cite{2022_Lin_channel_estimation}, a three-stage OMP algorithm was proposed to estimate IRS cascaded channels. This method sequentially estimated the angles of departure (AoDs) at the UE, the angles of arrival (AoAs) at the BS, and the cascaded UE-IRS-BS channel in each stage.
In \cite{2022_He_cascaded_channel_ANM}, an ANM-based channel estimation algorithm was proposed to sequentially estimate the angular parameters, angle differences, and the product of propagation path gains. In conventional IRS-assisted systems, the cascaded channel is typically estimated, under the assumption that the phase shifts of all elements can be dynamically adjusted. However, the situation differs in HE-IRS-assisted systems due to the coexistence of both DTEs and STEs. Given that the phase shifts of the STEs cannot be dynamically adjusted, the feasibility of estimating the overall cascaded channel still remains uncertain.
% Furthermore, exploring tailored channel estimation schemes by leveraging the characteristics of the HE-IRS structure is crucial for the green HE-IRS-assisted systems.
% not only reduces power consumption, but also has the potential to achieve pilot overhead reduction during channel estimation.

In this paper, we investigate the channel estimation and beamforming optimization for HE-IRS-assisted multi-user MIMO systems. By leveraging the characteristics of the HE-IRS structure, we propose tailored channel estimation and beamforming schemes to improve the sum rate performance. The main contributions are summarized below:
\begin{itemize}
% \item Unlike full cascaded channel estimation in conventional IRS, the HE-IRS channel to be estimated is decomposed into a DTE-based cascaded channel and an STE-based equivalent channel based on the unique DTE-STE integrated structure. The DTE-based cascaded channel is similar to those in conventional IRSs, with its dimension dependent to the number of DTEs, while the STE-based channel has a dimension independent of the number of STEs. With this decomposition, the dimension of the estimated HE-IRS channel reduces, thus lowering corresponding pilot overhead during channel estimation.
\item Based on the unique DTE-STE integrated structure, an operating process for HE-IRS-assisted multi-user MIMO systems is proposed. Before deploying the HE-IRS, the static phase shifts of the STEs are configured offline based on the prior information of the deployment locations of the BS and the HE-IRS, as well as the location area of the UEs. After deploying the HE-IRS, channel estimation is performed to acquire the HE-IRS channel, after which the phase shifts of the DTEs and the BS precoder are dynamically optimized according to the instantaneous CSI.
% the estimated HE-IRS channel and the pre-configured STE phase shifts are used to dynamically adjust the DTE phase shifts and the BS precoder.
\item In contrast to the overall cascaded channel estimated in conventional IRSs, we show that the HE-IRS channel to be estimated is decomposed into a DTE-based cascaded channel and an STE-based equivalent channel. By leveraging the DTE-STE decoupled channel model, the inherent sparsity of the DTE- and STE-based channels, and the manifold optimization (MO) method, we propose the channel estimation scheme. Three channel components in DTE- and STE-based channels are alternately estimated during outer and inner iterations. By considering scenarios with the rank mismatch problem in the imperfect channel sparsity information, we also develop a robust rank selection rule.
% \item Given the static phase-tuning capability of the STEs, we propose an offline beamforming strategy of de-focusing the reflect beam power to achieve broad coverage of the UE location area, and then propose a wide beam synthesis-MO (WBS-MO) optimization algorithm for the STEs. With the estimated DTE-based cascaded and STE-based equivalent channels, and the pre-determined phase shifts of the STEs, we propose a weighted mean square error minimization-element iteration (WMMSE-EI) algorithm to dynamically optimize the beamforming of the BS and the DTEs.
\item Given the coexistence of static and dynamic phase-shifting elements, the beamforming optimization for HE-IRS is divided into offline and online stages, corresponding to the STEs and the DTEs, respectively. In the offline stage, instead of targeting specific UEs, we propose a wide beam synthesis optimization algorithm for the STEs to concentrate the reflected beam power on the entire UE location area. In the online stage, by using the estimated DTE-based cascaded and STE-based equivalent channels, along with the pre-configured phase shifts of the STEs, we propose a beamforming algorithm to optimize the phase shifts of the DTEs and the BS precoder for specific UEs.
\item We provide various simulation results to verify the effectiveness
of the proposed channel estimation and beamforming schemes, and show that the green HE-IRS, with reduced pilot overhead, can achieve a competitive sum rate performance to that of the conventional IRS in multi-user MIMO systems, thanks to the efficient utilization of the unique DTE-STE integrated structure with the proposed channel estimation and beamforming schemes.
\end{itemize}

The rest of this paper is organized as follows: Section \ref{sec:Heterogeneous-IRS} introduces the structure of HE-IRS and its operating process. Section \ref{sec:System Model and Channel Model} presents the system model and channel model of the HE-IRS-assisted multi-user MIMO system. In Section \ref{sec:Channel Estimation Scheme}, the channel estimation scheme is proposed, along with a robust rank selection rule to address scenarios with channel rank mismatches. Section \ref{sec:Two-Stage Beamforming Optimization} proposes the offline beamforming algorithm for the STEs and the online beamforming algorithm for the BS and the DTEs. Simulation results are provided in Section \ref{sec:Simulations}. Finally, Section \ref{sec:Conclusion} draws the conclusions of this paper.

\emph{Notations:} $\mathrm{j}=\sqrt{-1}$ is the imaginary unit. $a$, $\mathbf{a}$ and $\mathbf{A}$ represent a scalar, a column vector and a matrix, respectively. $[\mathbf{a}]_i$ represents the $i$-th element of $\mathbf{a}$, and $[\mathbf{A}]_{ij}$ represents the $(i,j)$-th element of $\mathbf{A}$. $(\cdot)^T$, $(\cdot)^H$, $(\cdot)^*$, $\mathrm{Tr(\cdot)}$, $\mathrm{rank(\cdot)}$ and $\mathrm{vec(\cdot)}$ denote the transpose, conjugate transpose, conjugate operators, trace, rank and vectorization of a matrix, respectively. $\|\cdot\|$ denotes the Frobenius norm of a matrix, and $\|\cdot\|_N$ denotes the $\ell_N$-norm of a vector. $|\cdot|$ denotes the determinant (module) of a matrix (scalar). $\mathcal{R}\{\cdot\}$ denotes the real part of a scalar. $\mathbb{E}(\cdot)$ is the expectation operator. $\circ$, $\odot$ and $\otimes$ denotes Hadamard, Khatri-Rao and Kronecker products, respectively. $\mathrm{diag}(\mathbf{a})$ denotes a diagonal matrix with the elements of $\mathbf{a}$ on its main diagonal, and $\mathrm{diag}(\mathbf{A})$ is the extraction of the diagonal of $\mathbf{A}$. $\mathrm{blkdiag}(\mathbf{A}_1,\dots,\mathbf{A}_n)$ denotes a block diagonal matrix with diagonal components $\mathbf{A}_1,\dots,\mathbf{A}_n$. $\mathcal{CN}(\mathbf{0}, \mathbf{K})$ denotes the circularly symmetric complex Gaussian distribution with zero mean and covariance matrix $\mathbf{K}$.
\section{Heterogeneous-IRS}\label{sec:Heterogeneous-IRS}
In this section, we first introduce the structure of the HE-IRS and the inspiration behind its hardware design, and then introduce the whole operating process of HE-IRS-assisted systems. 
% based the HE an HE-IRS-assisted multi-user MIMO system model is established, followed by a detailed description of the channel model within this system.
\subsection{Heterogeneous-IRS Structure}\label{subsec:Heterogeneous-IRS}
\begin{figure}
	\centering
    \captionsetup{font=small}
	\includegraphics[width=0.6\columnwidth]{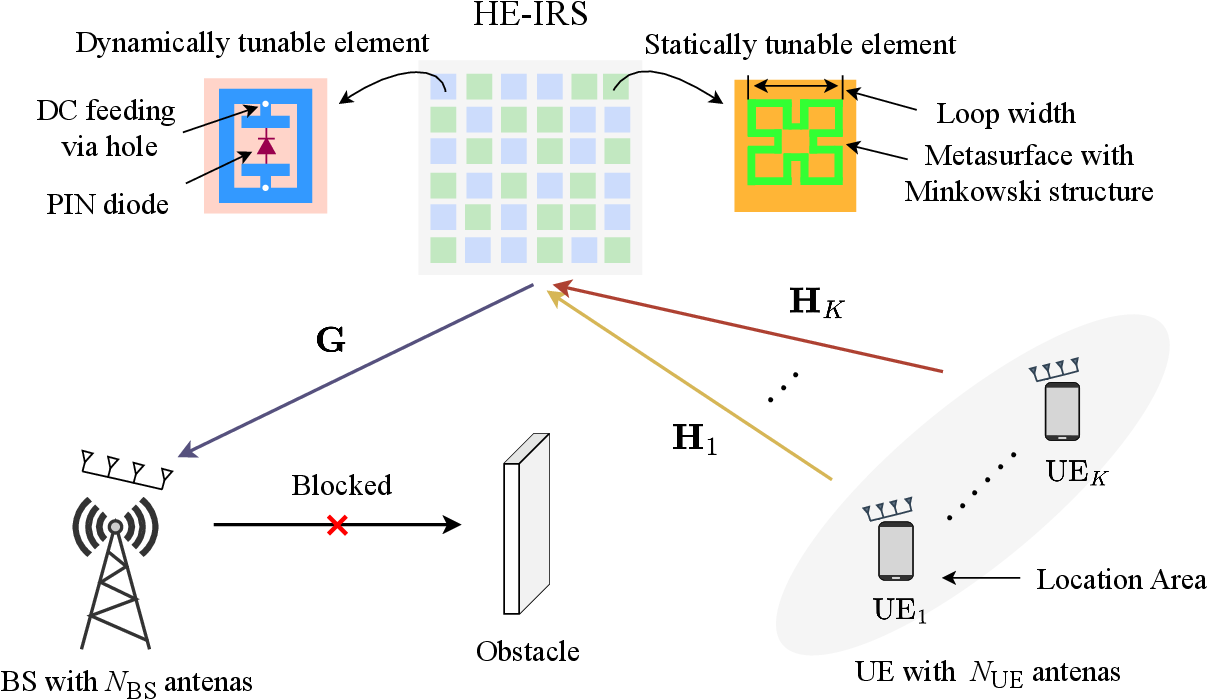}
	\caption{An HE-IRS-assisted multi-user MIMO system.}	\label{fig:model}
\end{figure}

As mentioned in Section \ref{sec:Introduction}, conventional IRS may have the issue of non-negligible power consumption and hardware complexity, especially when the system size becomes large. To overcome this issue, we have proposed an innovative HE-IRS in \cite{2024_Zhao_HEIRS}, as shown in Fig. \ref{fig:model}, which integrates two types of reflecting elements, i.e., DTEs and STEs. The DTEs possess the dynamic phase-tuning capability, similar to the reconfigurable elements used in conventional IRS, which can dynamically provide discrete phase shifts for incident signals by altering the states of PIN diodes \cite{2019_Wu_IRS_discrete}. However, such components, along with their corresponding control circuits, introduce additional power consumption and hardware complexity.
% However, the use of such components and corresponding control circuits introduces specific power consumption and hardware costs. 
As the number of DTEs scales to hundreds or even thousands, even achieving a 1-bit phase shift resolution can result in significant power consumption and hardware costs \cite{2024_Jin_Power}. Building on this observation, the STEs are designed
% further minimize phase shift resolution, effectively achieving a "0-bit" configuration, which means the STEs can only generate
with fixed phase shifts. This inherent static phase-tuning capability allows STEs to be implemented using pre-designed structures, thus eliminating the need for the components and control circuits required by DTEs \cite{2018_Optical_metasurfaces}. An example of an STE is shown in Fig. \ref{fig:model}, where its phase shift can be precisely determined by changing the loop width of the Minkowski structure \cite{2021_Minkowski_antenna}\cite{2017_Cui_Minkowski}. Since such pre-designed structures are entirely passive, the STEs consume no power. Consequently, the integration of power-free STEs significantly reduces the power consumption of the HE-IRS compared to conventional IRS with the same number of elements. Furthermore, the hardware complexity associated with HE-IRS is also reduced, as each STE no longer requires any control from the FPGA control board.
% Due to the unique DTE-STE integrated structure, the HE-IRS introduce a new collaborative way that sets it apart from conventional IRS. Specifically, when an incident signal impinges on the HE-IRS, two distinct types of reflected beams are generated: one resulting from the static yet precise phase shifts of the STEs, and the other from the dynamic but discrete phase shifts of the DTEs. As a result, these two types of reflected beams collaborate to synthesize the overall reflected beam of the HE-IRS, thereby enabling its operation within HE-IRS-assisted systems.
% \footnote{In the simulation, we will validate the collaboration between DTEs and STEs by examining the reflected beam patterns of the HE-IRS for the data stream of each UE.}
\subsection{HE-IRS Operating Process}\label{subsec:An Optimization Framework}
Due to the unique HE-IRS structure, in a typical HE-IRS-assisted system, the whole operating process may include the following steps:

\textit{Offline beamforming:} Before deploying the HE-IRS, an offline beamforming algorithm is needed to optimize the static phase shifts of the STEs according to where the BS and the HE-IRS are deployed and where the UEs are located.
% targeting the UE location area to achieve broad coverage.

\textit{Channel estimation:} After deploying the HE-IRS, the HE-IRS channel needs to be estimated by exploiting the characteristics of the unique DTE-STE integrated structure.

\textit{Online beamforming:} After channel estimation, an online beamforming algorithm is needed to optimize both the phase shifts of the DTEs and the BS precoder according to the instantaneous CSI.

% For clarity, channel estimation will be first discussed in Section IV, while offline and online beamforming will be discussed in Section V.

\section{System Model and Channel Model}\label{sec:System Model and Channel Model}
In this section, we introduce an HE-IRS-assisted multi-user MIMO system model, followed by a detailed description of the channel model within this system.
\subsection{System Model}\label{subsec:System Model}
We consider a narrowband multi-user MIMO system, as illustrated in Fig. \ref{fig:model}, where a BS equipped with $N_\mathrm{BS}$ antennas serves $K$ UEs, each equipped with $N_\mathrm{UE}$ antennas. To facilitate communication between the BS and the UEs, an HE-IRS with $N_\mathrm{DTE}$ DTEs and $N_\mathrm{STE}$ STEs $(N_\mathrm{DTE}+N_\mathrm{STE}=N)$ is deployed. The UEs are randomly located in an area covered by the HE-IRS. The uplink channel estimation is performed in the time division duplex (TDD) mode, and the downlink channels are obtained by estimating the uplink channels based on channel reciprocity. To simplify the estimation process, it is assumed that $K$ UEs transmit their pilots sequentially to the BS \cite{2024_Wu_BIOS_channel_estimation}. Taking the $k$-th UE as an example, the uplink baseband received signal at the BS is expressed as
\begin{equation} \label{eqn:received signal}
    \mathbf{r}_{k}[t]=\mathbf{G}\boldsymbol{\Psi}_{k}[t]\mathbf{H}_{k}\mathbf{s}_{k}[t]+\mathbf{z}_{k}[t],
\end{equation}
where $\mathbf{s}_{k}[t]\in\mathbb{C}^{N_{\mathrm{UE}}\times 1}$ denotes the $t$-th pilot vector satisfying the normalized power constraint $\big\|\mathbf{s}_{k}[t]\big\|^2=P_\mathrm{tr}$, and $\mathbf{z}_{k}[t]\in\mathbb{C}^{N_{\mathrm{BS}}\times 1}$ represents the additive Gaussian noise modeled as $\mathbf{z}_{k}[t]\sim\mathcal{CN}(0, \sigma^2\mathbf{I}_{N_{\mathrm{BS}}})$. $\boldsymbol{\Psi}_{k}[t]=\mathrm{diag}( \boldsymbol{\psi}_{k}[t] )\in \mathbb{C}^{N\times N}$ denotes the overall phase shift matrix of the HE-IRS. The HE-IRS is equipped with a uniform planar array (UPA) with a size of $N_\mathrm{y}\times N_\mathrm{z}$. For simplicity, we consider that both the DTEs and STEs are implemented as two adjacent
 UPAs with $N_\mathrm{DTE}$ $=N^{\mathrm{y}}_\mathrm{DTE}\times$ $ N_\mathrm{z}$ and $N_\mathrm{STE}$ $=N^{\mathrm{y}}_\mathrm{STE}\times$ $ N_\mathrm{z}$, where $N_\mathrm{DTE}+$ $N_\mathrm{STE}=$ $N_\mathrm{y}$\footnote{The proposed schemes are readily adaptable to the integrated implementation of DTEs and STEs.}. Accordingly, the phase shift vector of the HE-IRS can be expressed as  $\boldsymbol{\psi}_{k}[t]={\left[ \boldsymbol{\phi}_{k}^H[t],\boldsymbol{\omega}^H \right]}^H$, where $\boldsymbol{\phi}_{k}[t]\in \mathbb{C}^{N_\mathrm{DTE}\times 1}$ and $\boldsymbol{\omega}\in \mathbb{C}^{N_\mathrm{STE}\times 1}$ represent the phase shift vectors of the DTEs and STEs. $\mathbf{G}\in\mathbb{C}^{N_{\mathrm{BS}} \times N}$ represents the channel from the HE-IRS to the BS, and ${\mathbf{H}}_{k}\in\mathbb{C}^{N \times N_{\mathrm{UE}}}$ represents the channel between the $k$-th UE and the HE-IRS. In this work, the direct links between the BS and the UEs are assumed to be negligible due to severe blockages \cite{2022_Lin_channel_estimation}.

\subsection{Channel Model}\label{subsec:Channel Model}
We assume that the BS and the UEs are equipped with uniform linear array (ULA) antennas. By employing the Saleh-Valenzuela model \cite{2022_Lin_channel_estimation} to characterize the propagation environment, the HE-IRS to BS and the $k$-th UE to HE-IRS channel matrices can be expressed as
\begin{equation} \label{eqn:IRS-BS channel}
\mathbf{G}=\sqrt{\frac{N_{\mathrm{BS}}N}{P}}\sum\limits_{p=1}^P{\alpha_p \mathbf{a}_\mathrm{BS}\left( \theta_\mathrm{r}^p \right) \mathbf{a}_\mathrm{I}^H\left( \theta_\mathrm{t}^p, \varphi_\mathrm{t}^p \right)},
\end{equation}
\begin{equation} \label{eqn:UE-IRS channel}
\mathbf{H}_k=\sqrt{\frac{N_{\mathrm{UE}}N}{Q}}\sum\limits_{q=1}^Q{\beta_{q,k} \mathbf{a}_\mathrm{I}\left( \vartheta_{\mathrm{r},k}^q, \zeta_{\mathrm{r},k}^q \right)\mathbf{a}_\mathrm{UE}^H\left( \vartheta_{\mathrm{t},k}^q \right)},
\end{equation}
where $P$ and $Q$ denote the numbers of the paths of the IRS-BS and UE-IRS channels, respectively. Specifically, $p=1$ and $q=1$ correspond to the line-of-sight (LoS) paths, while $p=2,$$\dots$$,P$ and $q=2,\dots,Q$ correspond to the non-LoS (NLoS) paths. For the IRS-BS channel, ${\alpha }_p$, ${\theta}_\mathrm{r}^p$ and ${\theta}_\mathrm{t}^p$ (${\varphi}_\mathrm{t}^p$) denote the complex gain, the angle of arrival (AoA) and the azimuth (elevation) angles of departure (AoDs) of the $p$-th path. For the UE-IRS channel, ${\beta }_{q,k}$, ${\vartheta}_{\mathrm{r},k}^q$ (${\zeta}_{\mathrm{r},k}^q$) and ${\vartheta}_{\mathrm{t},k}^q$ denote the complex gain, azimuth (elevation) AoA and AoD of the $q$-th path. In addition, $\mathbf{a}_{\mathrm{BS}}$, $\mathbf{a}_{\mathrm{I}}$ and $\mathbf{a}_{\mathrm{UE}}$ represent the array response vectors at the BS, HE-IRS and UEs. Specifically, the ULA response vector is defined as
\begin{equation} \label{eqn:array vecor}
    \mathbf{a}\left( u,M \right)=\frac{1}{\sqrt{M}}{{\left[ 1,{{e}^{\mathrm{j}\pi u}},\dots,{{e}^{\mathrm{j}\pi \left( M-1 \right)u}} \right]}^{T}}.
\end{equation}
Accordingly, the array response vectors at the BS and UEs are given by $\mathbf{a}_\mathrm{BS}\left( \theta_\mathrm{r}^p \right)=\mathbf{a}( \cos ({\theta_\mathrm{r}^p}), {N}_{\mathrm{BS}})$ and $\mathbf{a}_\mathrm{UE}( \vartheta_{\mathrm{t},k}^q )=\mathbf{a}(\cos({\vartheta_{\mathrm{t},k}^q}), {N}_{\mathrm{UE}})$. For the HE-IRS, the array response vector is  ${{\mathbf{a}}_\mathrm{I}}$ $(\theta,\varphi)$ $=\mathbf{a}_y(\theta,\varphi) \otimes$ $ \mathbf{a}_z(\varphi)$, where $\mathbf{a}_y$ $(\theta,\varphi)=$ $\mathbf{a}$ $(\sin(\theta)$ $\sin(\varphi),$ $N_y)$ and $\mathbf{a}_z$ $(\varphi)=$ $\mathbf{a}(\cos(\varphi),N_z)$, with $\theta$ and $\varphi$ denoting the azimuth and elevation AoDs or AoAs of the HE-IRS. 

\section{DSD-MO Channel Estimation Scheme}\label{sec:Channel Estimation Scheme}
In this section, we first reveal the limitation that the conventional  overall cascaded channel estimation cannot be directly applied in HE-IRS-assisted systems. To tackle this issue, we propose a DTE-STE decoupled (DSD) model for the HE-IRS channel. By further exploiting the inherent sparsity of the DTE- and STE-based channels and applying the MO method, we design the DSD-MO scheme. 
% Following this, we formulate the corresponding channel estimation problem and solve it using the MV-MO algorithm.
We also develop a robust rank selection rule to tackle the scenarios with channel rank mismatches.
% By further considering the imperfect sparsity information with rank mismatches, a rank selection rule is developed, which ultimately forms the DSD-MO estimation scheme.
\subsection{The DTE-STE Decoupled Channel Model}\label{subsec:The Channel Estimation Challenge for HE-IRS}
In conventional IRS-assisted systems, the cascaded UE-IRS-BS channel is typically estimated \cite{2020_Yuan_cascaded_channel}. However, in HE-IRS-assisted systems, the fixed phase shifts introduced by the STEs poses a significant limitation, as \textit{the overall cascaded channel estimation cannot be directly applied}. In particular, we first express the received signal in (\ref{eqn:received signal}) as the cascaded channel form
\begin{equation} \label{eqn:received signal Hca}
     \mathbf{r}_{k}[t] = ({\boldsymbol{\psi}_k^T[t]}\otimes {\mathbf{s}_k^T[t]}\otimes {\mathbf{I}_{N_\mathrm{BS}}})\mathrm{vec}(\mathbf{H}_{\mathrm{ca},k})+\mathbf{z}_{k}[t],
\end{equation}
where $\mathbf{H}_{\mathrm{ca},k}=(\mathbf{H}_k^T \odot \mathbf{G})\in \mathbb{C}^{N_\mathrm{BS}N_\mathrm{UE}\times N}$ represents the cascaded channel of the HE-IRS for the $k$-th UE. The derivation of (\ref{eqn:received signal Hca}) follows from the facts that $\mathrm{vec}(\mathbf{ABD})=(\mathbf{D}^T\otimes\mathbf{A})\mathrm{vec}(\mathbf{B})$ and $\mathrm{vec}(\mathbf{ACD})=(\mathbf{D}^T\odot\mathbf{A})\mathbf{c}$ for any matrices $\mathbf{A}$, $\mathbf{B}$ and $\mathbf{D}$, and diagonal matrix $\mathbf{C}$ with $\mathbf{c}=\rm{diag}(\mathbf{C})$. To illustrate the limitation mentioned above, we consider the overall cascaded channel estimated by the conventional LS method \cite{2022_Jensen_RIS_LS}, as detailed in the following lemma.
\begin{lemma}\label{lem:Hca of HE-IRS}
    In the absence of noise effects in channel estimation, the conventional LS cascaded channel estimation fails to achieve perfect estimation of $\mathbf{H}_{\mathrm{ca},k}$ when $N_\mathrm{STE} > 1$.
\end{lemma}
\emph{Proof}: The received signal in (\ref{eqn:received signal Hca}) with no noise effect stacked over $T$ pilot transmissions is expressed as
\begin{equation} \label{eqn:received signal Hca T}
\underbrace{
\begin{bmatrix}
\mathbf{r}_k[1] \\
\mathbf{r}_k[2] \\
\vdots \\
\mathbf{r}_k[T]
\end{bmatrix}
}_{\mathbf{r}_k}
=
\underbrace{
\begin{bmatrix}
{\boldsymbol{\psi}_k^T[1]}\otimes {\mathbf{s}_k^T[1]}\otimes {\mathbf{I}_{N_\mathrm{BS}}} \\
{\boldsymbol{\psi}_k^T[2]}\otimes {\mathbf{s}_k^T[2]}\otimes {\mathbf{I}_{N_\mathrm{BS}}} \\
\vdots \\
{\boldsymbol{\psi}_k^T[T]}\otimes {\mathbf{s}_k^T[T]}\otimes {\mathbf{I}_{N_\mathrm{BS}}}
\end{bmatrix}
}_{\mathbf{J}_k}
\mathrm{vec}(\mathbf{H}_{\mathrm{ca},k})
% + \underbrace{
% \begin{bmatrix}
% \mathbf{z}_k[1] \\
% \mathbf{z}_k[2] \\
% \vdots \\
% \mathbf{z}_k[T]
% \end{bmatrix}
% }_{\mathbf{z}_k}
,
\end{equation}
where $\mathbf{r}_k \in \mathbb{C}^{N_\mathrm{BS}T \times 1}$ 
% and $\mathbf{z}_k \in \mathbb{C}^{N_\mathrm{BS}T \times 1}$ 
represents the stacked received signal, and $\mathbf{J}_k \in \mathbb{C}^{N_\mathrm{BS}T \times N_\mathrm{UE}N_\mathrm{BS}N}$ represents the equivalent measurement matrix constructed by stacking the terms ${\boldsymbol{\psi}_k^T[t]}\otimes {\mathbf{s}_k^T[t]}\otimes {\mathbf{I}_{N_\mathrm{BS}}}$ for $t=1,2,\dots,T$. The classic LS estimator is $\mathrm{vec}(\widehat{\mathbf{H}}_{\mathrm{ca},k})=\operatorname{argmin}_{\mathrm{vec}(\mathbf{H}_{\mathrm{ca},k})} \big\| \mathbf{r}_k - \mathbf{J}_k {\mathrm{vec}(\mathbf{H}_{\mathrm{ca},k})}\big\|^2$. It is clear that one crucial factor for achieving a perfect estimation of $\mathbf{H}_{\mathrm{ca},k}$ is the rank of the measurement matrix $\mathbf{J}_k$, which can be further expressed as
\begin{equation}\label{eqn:received signal Jk}
     \mathbf{J}_k = 
     \begin{bmatrix}
        {\boldsymbol{\psi}_k^T[1]}\otimes {\mathbf{s}_k^T[1]} \\
        {\boldsymbol{\psi}_k^T[2]}\otimes {\mathbf{s}_k^T[2]} \\
        \vdots \\
        {\boldsymbol{\psi}_k^T[T]}\otimes {\mathbf{s}_k^T[T]}
    \end{bmatrix}
    \otimes {\mathbf{I}_{N_\mathrm{BS}}} = (\mathbf{L}_k \odot \mathbf{S}_k)^T \otimes {\mathbf{I}_{N_\mathrm{BS}}},
\end{equation}
where $\mathbf{L}_k = \left[ \boldsymbol{\psi}_k[1],\dots,\boldsymbol{\psi}_k[T] \right] \in \mathbb{C}^{N \times T}$ and $\mathbf{S}_k = \left[ \mathbf{s}_k[1],\dots,\mathbf{s}_k[T] \right] \in \mathbb{C}^{N_\mathrm{UE} \times T}$. 
%(\ref{eqn:received signal Jk}) follows from the properties of $\otimes$ and $\odot$. 
Since the column space of $\mathbf{L}_k \odot \mathbf{S}_k$ is a subset of the column space of $\mathbf{L}_k \otimes \mathbf{S}_k$, i.e., $\operatorname{Col}(\mathbf{L}_k \odot \mathbf{S}_k) \subseteq \operatorname{Col}(\mathbf{L}_k \otimes \mathbf{S}_k)$, the rank of $\mathbf{L}_k \odot \mathbf{S}_k$ satisfies 
\begin{equation}\label{eqn:rank kro}
    \mathrm{rank}(\mathbf{L}_k \odot \mathbf{S}_k) \le \mathrm{rank}(\mathbf{L}_k \otimes \mathbf{S}_k) = \mathrm{rank}(\mathbf{L}_k) \cdot \mathrm{rank}(\mathbf{S}_k),
\end{equation}
where the derivation of (\ref{eqn:rank kro}) follows from the property  $\mathrm{rank}(\mathbf{A} \otimes \mathbf{B}) = \mathrm{rank}(\mathbf{A}) \cdot \mathrm{rank}(\mathbf{B})$ for any matrices $\mathbf{A}$ and $\mathbf{B}$. Next, we analyze the ranks of $\mathbf{L}_k$ and $\mathbf{S}_k$. For $\mathbf{L}_k$, if $N_\mathrm{STE} > 1$, there exist $N_\mathrm{STE}$ rows in the matrix where all elements within each individual row are identical across all $T$ columns, i.e., $[\mathbf{L}_k]_{l1}=[\mathbf{L}_k]_{l2}=\dots=[\mathbf{L}_k]_{lT}$ for $l \in \{ 1,\dots,N \}$. This structural property introduces linear dependence among these $N_\mathrm{STE}$ rows, thereby reducing the rank of $\mathbf{L}_k$. As a result, $\mathrm{rank}(\mathbf{L}_k) < N$. For $\mathbf{S}_k$, its rank is limited by the row dimension, i.e., $\mathrm{rank}(\mathbf{S}_k) \le N_\mathrm{UE}$. Thus, if $N_\mathrm{STE} > 1$, it follows that $\mathrm{rank}(\mathbf{L}_k) \cdot \mathrm{rank}(\mathbf{S}_k) < N_\mathrm{UE} N$. Given that the Kronecker product with ${\mathbf{I}_{N_\mathrm{BS}}}$ further scales the rank by ${N_\mathrm{BS}}$, we have 
\begin{equation}\label{eqn:Jk rank}
    \mathrm{rank}(\mathbf{J}_k) = \mathrm{rank}((\mathbf{L}_k \odot \mathbf{S}_k)^T \otimes {\mathbf{I}_{N_\mathrm{BS}}}) < N_\mathrm{UE}N_\mathrm{BS}N.
\end{equation}
Hence, $\mathbf{J}_k$ is column rank-deficient when $N_\mathrm{STE} > 1$, and consequently, the cascaded channel $\mathbf{H}_{\mathrm{ca},k}$ cannot be perfectly estimated. This completes the proof.$\hfill\blacksquare$

As observed in Lemma \ref{lem:Hca of HE-IRS}, these fixed phase shifts of the STEs induce structural dependencies in the measurement matrix used for estimating the HE-IRS cascaded channel, thereby making perfect estimation infeasible. As a result, the overall cascaded channel estimation cannot be directly applied for HE-IRS channel estimation. To overcome this limitation, we propose to decouple the HE-IRS channel into two separate components, corresponding to the DTEs and the STEs. Based on this, (\ref{eqn:received signal}) is reformulated as
\begin{equation} \label{eqn:received signal sep}
    \mathbf{r}_{k}[t]=(\underbrace{\mathbf{G}_\mathrm{DTE}\boldsymbol{\Phi}_{k}[t]\mathbf{H}_{\mathrm{DTE},k}}_{\text{DTE-based channel}} + \underbrace{\mathbf{G}_\mathrm{STE}\boldsymbol{\Omega}\mathbf{H}_{\mathrm{STE},k}}_{\text{STE-based channel}} )\mathbf{s}_{k}[t]+\mathbf{z}_{k}[t],
\end{equation}
where $\boldsymbol{\Phi}_{k}[t]=\mathrm{diag}(\boldsymbol{\phi}_{k}[t]) \in \mathbb{C}^{N_\mathrm{DTE} \times N_\mathrm{DTE}}$ is the uplink phase shift matrix of the DTEs for the $t$-th pilot vector of the $k$-th UE, while $\boldsymbol{\Omega} = \mathrm{diag}(\boldsymbol{\omega}) \in \mathbb{C}^{N_\mathrm{STE} \times N_\mathrm{STE}}$ is the phase shift matrix of the STEs, which remains constant across all pilots and UEs. $\mathbf{G}_\mathrm{DTE} \in \mathbb{C}^{N_\mathrm{BS} \times N_\mathrm{DTE}}$ ($\mathbf{G}_\mathrm{STE} \in \mathbb{C}^{N_\mathrm{BS} \times N_\mathrm{STE}}$) denotes the HE-IRS to BS channel corresponding to the DTEs (STEs), and $\mathbf{H}_{\mathrm{DTE},k}  \in \mathbb{C}^{N_\mathrm{DTE} \times N_\mathrm{UE}}$ ($\mathbf{H}_{\mathrm{STE},k} \in \mathbb{C}^{N_\mathrm{STE} \times N_\mathrm{UE}}$) represents the $k$-th UE to HE-IRS channel corresponding to the DTEs (STEs). 

% $\mathbf{H}_{\mathrm{ca},k}^{\mathrm{DTE}}= \mathbf{H}_{\mathrm{DTE},k}\odot \mathbf{G}_{\mathrm{DTE}}$

From (\ref{eqn:received signal sep}), the overall HE-IRS channel is reformulated as a  DTE-STE decoupled channel. Specifically, for the DTE-based channel, since $\boldsymbol{\Phi}_{k}[t]$ needs to be dynamically adjusted during beamforming, the cascaded channel of $\mathbf{G}_{\mathrm{DTE}}$ and $\mathbf{H}_{\mathrm{DTE},k}$ must be estimated. In contrast, for the STE-based channel, the equivalent channel
$\mathbf{H}_{\mathrm{eq},k}^{\mathrm{STE}}= \mathbf{G}_\mathrm{STE}\boldsymbol{\Omega}\mathbf{H}_{\mathrm{STE},k}\in \mathbb{C}^{N_\mathrm{BS}\times{N_\mathrm{UE}}}$ can be directly estimated as a single entity, as $\boldsymbol{\Omega}$ is fixed and does not require online optimization. The channel representation in (\ref{eqn:received signal sep}) addresses the limitation of overall cascaded channel estimation, which arises from the limited degrees of freedom due to the fixed phase shifts of the STEs.
% By alternately estimating $\mathbf{G}_{\mathrm{DTE}}$ and $\mathbf{H}_{\mathrm{DTE},k}$ corresponding to the DTEs and $\mathbf{H}_{\mathrm{eq},k}^{\mathrm{STE}}$ corresponding to the STEs within the DTE-STE decoupled model, the HE-IRS channel can be effectively estimated.

\textit{Remark 1:} Comparing the system assisted by the HE-IRS with that assisted by the conventional IRS with the same number of elements, we can see that the conventional IRS channel to be estimated is the overall cascaded channel, with a matrix dimension of $NN_\mathrm{BS}N_\mathrm{UE}$, while the HE-IRS channel to be estimated consists of the DTE-based cascaded channel and the STE-based equivalent channel, with a total matrix dimension of $(N_\mathrm{DTE}+1)N_\mathrm{BS}N_\mathrm{UE}$. It is observed that the matrix dimension of the HE-IRS channel is related to the number of the DTEs, $N_\mathrm{DTE}$, instead of the total number of elements, $N$, in both HE-IRS and conventional IRS. This implies that less pilot overhead is required in HE-IRS.  
% Besides, it is independent on the number of STEs, and only increasing the number of STEs does not affect the total variables requiring estimation. 
% Furthermore, define the ratio of variables to be estimated in the HE-IRS channel to that in the IRS channel as $\tau=(N_\mathrm{DTE}+1)/N$. Given that $N_\mathrm{DTE}$ is generally much larger than 1, $\tau$ is approximately equal to the ratio of the number of DTEs to the total number of elements.
% As can be observed in Lemma \ref{lem:Hca of HE-IRS}, these fixed phase shifts of the STEs induce structural dependencies in the measurement matrix used for estimating the cascaded channel of the HE-IRS, thereby making perfect estimation infeasible. As a result, typical cascaded channel estimation methods for conventional IRSs, whether based on matrix factorization or decomposition techniques \cite{2020_Yuan_cascaded_channel}\cite{2021_Araújo_cascaded_channel}\cite{2020_Liu_cascaded_channel}, or compressing sensing approaches \cite{2020_Wang_cascaded_channel_CS}\cite{2021_Mirza_cascaded_channel_CS}\cite{2022_He_cascaded_channel_ANM}, fail to address the unique channel estimation challenges in HE-IRS-assisted systems, as accurately estimating the overall cascaded channel of the HE-IRS is infeasible. To overcome this limitation and obtain more precise CSI for beamforming, a DTE-STE decoupled channel model is proposed in the following.
\subsection{The Sparsity of the HE-IRS Channel}\label{subsec:The DTE-STE Decoupled Channel and its Sparsity}
As mentioned in Section \ref{subsec:Channel Model}, the UE-IRS-BS links exhibit limited propagation paths due to significant path loss and blocking effects. Thus, similar to that in \cite{2023_yuwei_sparsity_channel}, the channels from the HE-IRS to the BS and from the $k$-th UE to the HE-IRS, corresponding to the DTEs and STEs, can be represented in an angular sparse format as
\begin{align}
    \mathbf{G}_\mathrm{DTE} & \approx \mathbf{A}_{\mathrm{BS}}\mathbf{\Lambda}_{\mathbf{G}_\mathrm{DTE}}\mathbf{A}_{\mathrm{DTE}}^H, & \mathbf{H}_{\mathrm{DTE},k} & \approx \mathbf{A}_{\mathrm{DTE}}\mathbf{\Lambda}_{\mathrm{H}_{\mathrm{DTE},k}}\mathbf{A}_{\mathrm{UE}}^H, \label{eqn:sparsity representation of DTE-basd channel} \\
    \mathbf{G}_\mathrm{STE} & \approx \mathbf{A}_{\mathrm{BS}}\mathbf{\Lambda}_{\mathbf{G}_\mathrm{STE}}\mathbf{A}_{\mathrm{STE}}^H, & \mathbf{H}_{\mathrm{STE},k} & \approx \mathbf{A}_{\mathrm{STE}}\mathbf{\Lambda}_{\mathrm{H}_{\mathrm{STE},k}}\mathbf{A}_{\mathrm{UE}}^H, \label{eqn:sparsity representation of STE-basd channel}
\end{align}
where $\mathbf{A}_{\mathrm{BS}}\in\mathbb{C}^{N_{\mathrm{BS}}\times G_{\mathrm{BS}}}$ and $\mathbf{A}_{\mathrm{UE}}\in\mathbb{C}^{N_{\mathrm{UE}}\times G_{\mathrm{UE}}}$ are the overcomplete angular domain dictionaries at the BS and UE, respectively, with angle resolutions of $G_{\mathrm{BS}}$ and $G_{\mathrm{UE}}$. Similarly, $\mathbf{A}_{\mathrm{DTE}}\in\mathbb{C}^{N_\mathrm{DTE} \times G_{\mathrm{DTE}}}$ and $\mathbf{A}_{\mathrm{STE}}\in\mathbb{C}^{N_\mathrm{STE} \times G_{\mathrm{STE}}}$ represent the dictionaries at the DTEs and STEs, with resolutions $G_{\mathrm{DTE}}$ and $G_{\mathrm{STE}}$. Each column of these dictionaries corresponds to a specific AoA/AoD at the BS, DTEs, STEs or the $k$-th UE. $\mathbf{\Lambda}_{\mathbf{G}_\mathrm{DTE}}\in\mathbb{C}^{G_{\mathrm{BS}}\times G_{\mathrm{DTE}}}$ ($\mathbf{\Lambda}_{\mathbf{G}_\mathrm{STE}}\in\mathbb{C}^{G_{\mathrm{BS}}\times G_{\mathrm{STE}}}$) and $\mathbf{\Lambda}_{\mathrm{H}_{\mathrm{DTE}, k}}\in\mathbb{C}^{G_{\mathrm{DTE}}\times G_{\mathrm{UE}}}$ ($\mathbf{\Lambda}_{\mathrm{H}_{\mathrm{STE}, k}}\in\mathbb{C}^{G_{\mathrm{STE}}\times G_{\mathrm{UE}}}$) are the angular domain sparse matrices for $\mathbf{G}_\mathrm{DTE}$ ($\mathbf{G}_\mathrm{STE}$) and $\mathbf{H}_{\mathrm{DTE},k}$ ($\mathbf{H}_{\mathrm{STE},k}$), containing $P$ and $Q$ non-zero elements corresponding to the channel path gains, respectively. 
% Similarly, $\mathbf{\Lambda}_{\mathbf{G}_\mathrm{STE}}\in\mathbb{C}^{G_{\mathrm{BS}}\times G_{\mathrm{STE}}}$ and $\mathbf{\Lambda}_{\mathrm{H}_{\mathrm{STE}, k}}\in\mathbb{C}^{G_{\mathrm{STE}}\times G_{\mathrm{UE}}}$ represent the sparse matrices for $\mathbf{G}_\mathrm{STE}$ and $\mathbf{H}_{\mathrm{STE},k}$, also containing $P$ and $Q$ non-zero elements corresponding to the channel path gains. 
By choosing codewords from a uniform grid, $\mathbf{A}_{\mathrm{BS}}$ and $\mathbf{A}_{\mathrm{UE}}$ can be expressed as
\begin{align}
    \mathbf{A}_{\mathrm{BS}} & = \left[\mathbf{a}\big(u^1_\mathrm{BS}, N_{\mathrm{BS}}\big),\dots,\mathbf{a}\big(u^{G_{\mathrm{BS}}}_\mathrm{BS}, N_{\mathrm{BS}}\big)\right], \mathbf{A}_{\mathrm{UE}} = \left[\mathbf{a}\big(u^1_\mathrm{UE}, N_{\mathrm{UE}}\big),\dots,\mathbf{a}\big(u^{G_{\mathrm{UE}}}_\mathrm{UE}, N_{\mathrm{UE}}\big)\right], \label{eqn:sparse respresentation of BS}
    % \mathbf{A}_{\mathrm{UE}} & = \left[\mathbf{a}\big(u^1_\mathrm{UE}, N_{\mathrm{UE}}\big),\dots,\mathbf{a}\big(u^{G_{\mathrm{UE}}}_\mathrm{UE}, N_{\mathrm{UE}}\big)\right], \label{eqn:sparse respresentation of UE}
\end{align}
where $u^i_\mathrm{BS} = -1 + (i-1)\frac{2}{G_\mathrm{BS}}$ and $u^j_\mathrm{UE} = -1 + (j-1)\frac{2}{G_\mathrm{UE}}$. Let $G_\mathrm{DTE}^\mathrm{y}$ and $G_\mathrm{DTE}^\mathrm{z}$ represent the angular resolutions of the DTEs along the y-axis and z-axis, $\mathbf{A}_{\mathrm{DTE}}$ can be expressed as $\mathbf{A}_{\mathrm{DTE}} = \mathbf{A}_{\mathrm{DTE}}^\mathrm{y}\otimes\mathbf{A}_{\mathrm{DTE}}^\mathrm{z}$, where $\mathbf{A}_{\mathrm{DTE}}^\mathrm{y} $ $= [\mathbf{a}(u^1_\mathrm{DTE,y},$ $ N_{\mathrm{DTE}}^\mathrm{y}),$ $\dots,\mathbf{a}(u^{G_\mathrm{y}}_\mathrm{DTE,y}$ $, N_{\mathrm{DTE}}^\mathrm{y})], $ $\mathbf{A}_{\mathrm{DTE}}^\mathrm{z}$ $ = [\mathbf{a}(u^1_\mathrm{DTE,z},$ $N_\mathrm{z}),$ $\dots,\mathbf{a}(u^{G_\mathrm{z}}_\mathrm{DTE,z}$ $,N_\mathrm{z})]$, with $u^{i}_\mathrm{DTE,y}$ $ = -1 + (i-1)\frac{2}{G_\mathrm{DTE}^\mathrm{y}}$, $u^{j}_\mathrm{DTE,z}$ $ = -1 + (j-1)\frac{2}{G_\mathrm{DTE}^\mathrm{z}}$, and $G_{\mathrm{DTE}}^\mathrm{y}G_{\mathrm{DTE}}^\mathrm{z} = G_{\mathrm{DTE}}$. Similarly,  $\mathbf{A}_{\mathrm{STE}}$ $ = \mathbf{A}_{\mathrm{STE}}^\mathrm{y}$ $\otimes\mathbf{A}_{\mathrm{STE}}^\mathrm{z}$, where $\mathbf{A}_{\mathrm{STE}}^\mathrm{y}$ $ = [{{e}^{\mathrm{j}\pi (N_\mathrm{DTE}^\mathrm{y}-1)u^1_\mathrm{STE,y}}}$ $, \dots, {{e}^{\mathrm{j}\pi (N_\mathrm{DTE}^\mathrm{y}-1)u^{G_\mathrm{y}}_\mathrm{STE,y}}}]$ $ \odot [\mathbf{a}(u^1_\mathrm{STE,y},$ $ N_{\mathrm{STE}}^\mathrm{y}),\dots,$ $\mathbf{a}(u^{G_\mathrm{y}}_\mathrm{STE,y},$ $ N_{\mathrm{STE}}^\mathrm{y})]$ and $
\mathbf{A}_{\mathrm{STE}}^\mathrm{z}$ $ = [\mathbf{a}(u^1_\mathrm{STE,z},N_\mathrm{z}),$ $\dots,\mathbf{a}(u^{G_\mathrm{z}}_\mathrm{STE,z},$ $N_\mathrm{z})]$ with $u^{i}_\mathrm{STE,y}$ $ = -1 + (i-1)\frac{2}{G_\mathrm{STE}^\mathrm{y}}$, $u^{j}_\mathrm{STE,z}$ $ = -1 + (j-1)\frac{2}{G_\mathrm{STE}^\mathrm{z}}$, and $G_{\mathrm{STE}}^\mathrm{y}$ $G_{\mathrm{STE}}^\mathrm{z} = G_{\mathrm{STE}}$.

Due to this sparsity of channels in the HE-IRS-assisted system, for the DTE- and STE-based channels, we can obtain the following two properties if the AoAs (AoDs) at the BS, HE-IRS, and UE are different from each other.
\begin{enumerate}
    \item For $\mathbf{G}_\mathrm{DTE}$ and $\mathbf{H}_{\mathrm{DTE},k}$ of the DTE-based channel:
    \begin{enumerate}
        \item \textbf{Angle sparsity:} If $G_{\mathrm{UE}}=N_{\mathrm{UE}}$, $G_{\mathrm{BS}}=N_{\mathrm{BS}}$ and $G_{\mathrm{DTE}}=N_{\mathrm{DTE}}$, we asymptotically have
        % \footnote{The angle sparsity is strictly applicable only when the AoAs and AoDs are on the grids of overcomplete angular domain dictionaries $\mathbf{A}_{\mathrm{BS}}$, $\mathbf{A}_{\mathrm{DTE}}$, and $\mathbf{A}_{\mathrm{UE}}$. 
        % As the resolutions $G_{\mathrm{UE}}, G_{\mathrm{BS}}, G_{\mathrm{DTE}} \to \infty$, (\ref{eqn:angle sparsity of DTE-based channel}) is asymptotically satisfied. However, in cases where these resolutions are not sufficiently high, our simulation suggests that exploiting this property in practical systems can still improve the channel estimation performance.} 
        \cite{2022_Lin_channel_estimation}
            \begin{equation}\label{eqn:angle sparsity of DTE-based channel}
                \left\|\boldsymbol{\lambda}_{\mathbf{G}_\mathrm{DTE}}\right\|_{0}=P,\ \left\|\boldsymbol{\lambda}_{\mathbf{H}_{\mathrm{DTE},k}}\right\|_{0}=Q,
            \end{equation}
            with $\boldsymbol{\lambda}_{\mathbf{G}_\mathrm{DTE}}=\mathrm{vec}(\mathbf{A}_{\mathrm{BS}}^H{\mathbf{G}_\mathrm{DTE}}\mathbf{A}_{\mathrm{DTE}})$ and $\boldsymbol{\lambda}_{\mathbf{H}_{\mathrm{DTE},k}}=\mathrm{vec}(\mathbf{A}_{\mathrm{DTE}}^H{\mathbf{H}_{\mathrm{DTE},k}}\mathbf{A}_{\mathrm{UE}})$.
    \item \textbf{Low-rank:} If $ \mathrm{min}\{N_{\mathrm{BS}},N_\mathrm{DTE}^\mathrm{y},N_\mathrm{DTE}^\mathrm{z}\}\ge P$ and $\mathrm{min}\{N_{\mathrm{UE}},N_\mathrm{DTE}^\mathrm{y},N_\mathrm{DTE}^\mathrm{z}\}\ge Q$, we have \cite{2022_Liu_ICC_2022}
        \begin{equation}\label{eqn:rank of DTE-based channel}
            \mathrm{rank}({\mathbf{G}_\mathrm{DTE}}) = P,\ \mathrm{rank}({\mathbf{H}_{\mathrm{DTE},k}}) = Q.
        \end{equation}
    \end{enumerate}
    \item For ${\mathbf{H}_{\mathrm{eq},k}^{\mathrm{STE}}}$ of the STE-based channel:
    \begin{enumerate}
        \item \textbf{Angle sparsity:} If $G_{\mathrm{UE}}=N_{\mathrm{UE}}$ and $G_{\mathrm{BS}}=N_{\mathrm{BS}}$, we asymptotically have
        \begin{equation}\label{eqn:angle sparsity of STE-based channel}
                \big\|{\boldsymbol{\Lambda}_{{\mathbf{H}_{\mathrm{eq},k}^{\mathrm{STE}}}}}\mathbf{1}_{N_\mathrm{UE}} \big\|_{0} \le P,\  \big\|\mathbf{1}_{N_\mathrm{BS}}^T{\boldsymbol{\Lambda}_{{\mathbf{H}_{\mathrm{eq},k}^{\mathrm{STE}}}}} \big\|_{0} \le Q,
            \end{equation}
            with $\boldsymbol{\Lambda}_{{\mathbf{H}_{\mathrm{eq},k}^{\mathrm{STE}}}}=\mathbf{A}_{\mathrm{BS}}^H{\mathbf{H}_{\mathrm{eq},k}^{\mathrm{STE}}}\mathbf{A}_{\mathrm{UE}}$.\\
            \textit{Proof:} Notice that when $G_{\mathrm{UE}}=N_{\mathrm{UE}}$ and  $G_{\mathrm{BS}}=N_{\mathrm{BS}}$, the dictionaries $\mathbf{A}_{\mathrm{BS}}$ and $\mathbf{A}_{\mathrm{UE}}$ are all unitary matrices according to (\ref{eqn:sparse respresentation of BS}). Thus, we have $\boldsymbol{\Lambda}_{{\mathbf{H}_{\mathrm{eq},k}^{\mathrm{STE}}}}=\mathbf{\Lambda}_{\mathbf{G}_\mathrm{STE}}\mathbf{A}_{\mathrm{STE}}^H\boldsymbol{\Omega}\mathbf{A}_{\mathrm{STE}}\mathbf{\Lambda}_{\mathrm{H}_{\mathrm{STE},k}}$. Since $\mathbf{\Lambda}_{\mathbf{G}_\mathrm{STE}}$ and $\mathbf{\Lambda}_{\mathrm{H}_{\mathrm{STE},k}}$ are sparse matrices with $P$ and $Q$ non-zero elements, respectively, it follows that $\boldsymbol{\Lambda}_{{\mathbf{H}_{\mathrm{eq},k}^{\mathrm{STE}}}}$ is a row-column-sparse matrix with at most $P$ non-zero rows and $Q$ non-zero columns. This completes the proof.$\hfill\blacksquare$
        \item \textbf{Low-rank:} If $ \mathrm{min}\{N_{\mathrm{BS}},N_\mathrm{STE}^\mathrm{y},N_\mathrm{STE}^\mathrm{z}\}\ge P$ and $\mathrm{min}\{N_{\mathrm{UE}},N_\mathrm{STE}^\mathrm{y},N_\mathrm{STE}^\mathrm{z}\}\ge Q$, then we have
        \begin{equation}\label{eqn:rank of STE-based channel}
            \mathrm{rank}({\mathbf{H}_{\mathrm{eq},k}^{\mathrm{STE}}}) \le \mathrm{min}\{P,Q\},
        \end{equation}
        \textit{Proof:} Similar to (\ref{eqn:rank of DTE-based channel}), we have $\mathrm{rank}({\mathbf{G}_\mathrm{STE}}) = P$ and $\mathrm{rank}({\mathbf{H}_{\mathrm{STE},k}}) = Q$. Given  $\mathrm{rank}(\mathbf{A} \mathbf{B}) \le \mathrm{min}\{\mathrm{rank}(\mathbf{A}),\mathrm{rank}(\mathbf{B})\}$ for any matrices $\mathbf{A}$ and $\mathbf{B}$, the proof is thus complete.$\hfill\blacksquare$
    \end{enumerate} 
\end{enumerate}

\textit{Remark 2:} For the equivalent matrix ${\mathbf{H}_{\mathrm{eq},k}^{\mathrm{STE}}}=\mathbf{G}_\mathrm{STE}\boldsymbol{\Omega}\mathbf{H}_{\mathrm{STE},k}$, its sparsity is influenced by a variable $\boldsymbol{\Omega}$ to be optimized. Consequently, the rank of ${\mathbf{H}_{\mathrm{eq},k}^{\mathrm{STE}}}$ and the numbers of the sparse rows and columns in ${\boldsymbol{\Lambda}_{{\mathbf{H}_{\mathrm{eq},k}^{\mathrm{STE}}}}}$ cannot always be mathematically guaranteed to achieve their respective maximum values of $\mathrm{min}\{P,Q\}$, $P$, and $Q$, unlike the cases of $\mathbf{G}_\mathrm{DTE}$ and $\mathbf{H}_{\mathrm{DTE},k}$ in the DTE-based channel. However, we observed from simulation tests that these metrics generally reach their expected maximum values. As a result, the fixed rank and the fixed numbers of sparse rows and columns will be utilized in the subsequent problem formulation and optimization algorithm. 
% Furthermore, cases where these metrics do not attain the fixed maximum values will be discussed in detail in the following chapter on robust channel estimation.

\subsection{Problem Formulation of Channel Estimation}\label{subsec:Problem Formulation of Channel Estimation}
Based on the system model presented in Section \ref{subsec:System Model}, the BS can estimate the channels corresponding to all UEs separately over consecutive time intervals. Without loss of generality, we focus on the estimation of the channels for the $k$-th UE, utilizing $T$ uplink pilot vectors. Besides, with the DTE-STE decoupled channel model discussed in Section \ref{subsec:The Channel Estimation Challenge for HE-IRS}, and the properties of angle sparsity and low-rank mentioned in Section \ref{subsec:The DTE-STE Decoupled Channel and its Sparsity}, the estimation problem for $\mathbf{G}_\mathrm{DTE}$, $\mathbf{H}_{\mathrm{DTE},k}$, and ${\mathbf{H}_{\mathrm{eq},k}^{\mathrm{STE}}}$ can be formulated as
\begin{equation}\label{eqn:LS estimation}
\begin{array}{cl}
\underset{\widehat{\mathbf{G}}_\mathrm{DTE}, \widehat{\mathbf{H}}_{\mathrm{DTE},k}, \widehat{\mathbf{H}}_{\mathrm{eq},k}^{\mathrm{STE}}}{\operatorname{minimize}} & \sum\limits_{t=1}^{T}\Big\|\mathbf{r}_{k}[t]-\widehat{\mathbf{G}}_\mathrm{DTE}\boldsymbol{\Phi}_{k}[t]\widehat{\mathbf{H}}_{\mathrm{DTE},k} \mathbf{s}_{k}[t]-\widehat{\mathbf{H}}_{\mathrm{eq},k}^{\mathrm{STE}}\mathbf{s}_{k}[t]\Big\|^{2} \\
\text { subject to } & \operatorname{rank}(\widehat{\mathbf{G}}_\mathrm{DTE})=P, \operatorname{rank}(\widehat{\mathbf{H}}_{\mathrm{DTE},k})=Q, \operatorname{rank}(\widehat{\mathbf{H}}_{\mathrm{eq},k}^{\mathrm{STE}})=\mathrm{min}\{P,Q\},\\
& \big\|\boldsymbol{\lambda}_{\widehat{\mathbf{G}}_\mathrm{DTE}}\big\|_{0}=P, \big\|\boldsymbol{\lambda}_{\widehat{\mathbf{H}}_{\mathrm{DTE},k}}\big\|_{0}=Q, \big\|{\boldsymbol{\Lambda}_{{\widehat{\mathbf{H}}_{\mathrm{eq},k}^{\mathrm{STE}}}}}\mathbf{1}_{N_\mathrm{UE}}\big\|_{0} = P, \big\|\mathbf{1}_{N_\mathrm{BS}}^T{\boldsymbol{\Lambda}_{{\widehat{\mathbf{H}}_{\mathrm{eq},k}^{\mathrm{STE}}}}}\big\|_{0} = Q.
\end{array}
\end{equation}
Unfortunately, problem (\ref{eqn:LS estimation}) is difficult to solve due to the highly non-convex constraints and the multiple coupled variables. Consequently, achieving a globally optimal solution may not be possible. To make this problem more tractable, we apply the $\ell_{1}$-norm regularization to relax the $\ell_{0}$-norm constrains \cite{2010_Berger_l1_norm}. Therefore, (\ref{eqn:LS estimation}) can be rewritten as
\begin{equation}\label{eqn:LS estimation l1 norm}
\begin{array}{cl}
\underset{\widehat{\mathbf{G}}_\mathrm{DTE}, \widehat{\mathbf{H}}_{\mathrm{DTE},k}, \widehat{\mathbf{H}}_{\mathrm{eq},k}^{\mathrm{STE}}}{\operatorname{minimize}} & \sum\limits_{t=1}^{T}\big\|\mathbf{r}_{k}[t]-\widehat{\mathbf{G}}_\mathrm{DTE}\boldsymbol{\Phi}_{k}[t]\widehat{\mathbf{H}}_{\mathrm{DTE},k}\mathbf{s}_{k}[t]-\widehat{\mathbf{H}}_{\mathrm{eq},k}^{\mathrm{STE}}\mathbf{s}_{k}[t]\big\|^{2} +\upsilon_{\widehat{\mathbf{G}}_\mathrm{DTE}}\big\|\boldsymbol{\lambda}_{\widehat{\mathbf{G}}_\mathrm{DTE}}\big\|_{1}\\
&+\upsilon_{\widehat{\mathbf{H}}_{\mathrm{DTE},k}}\big\|\boldsymbol{\lambda}_{\widehat{\mathbf{H}}_{\mathrm{DTE},k}}\big\|_{1}+\upsilon_{\widehat{\mathbf{H}}_{\mathrm{eq},k}^{\mathrm{STE}}}^{\mathrm{row}}\big\|{\boldsymbol{\Lambda}_{{\widehat{\mathbf{H}}_{\mathrm{eq},k}^{\mathrm{STE}}}}}\mathbf{1}_{N_\mathrm{UE}}\big\|_{1}+\upsilon_{{\widehat{\mathbf{H}}_{\mathrm{eq},k}^{\mathrm{STE}}}}^{\mathrm{col}}\big\|\mathbf{1}_{N_\mathrm{BS}}^T{\boldsymbol{\Lambda}_{{\widehat{\mathbf{H}}_{\mathrm{eq},k}^{\mathrm{STE}}}}}\big\|_{1}\\
\text { subject to } & \operatorname{rank}(\widehat{\mathbf{G}}_\mathrm{DTE})=P, \operatorname{rank}(\widehat{\mathbf{H}}_{\mathrm{DTE},k})=Q, \operatorname{rank}(\widehat{\mathbf{H}}_{\mathrm{eq},k}^{\mathrm{STE}})=\mathrm{min}\{P,Q\},\\
\end{array}
\end{equation}
where $\upsilon_{\widehat{\mathbf{G}}_\mathrm{DTE}}$, $\upsilon_{\widehat{\mathbf{H}}_{\mathrm{DTE},k}}$, $\upsilon_{\widehat{\mathbf{H}}_{\mathrm{eq},k}^{\mathrm{STE}}}^{\mathrm{row}}$, and $\upsilon_{{\widehat{\mathbf{H}}_{\mathrm{eq},k}^{\mathrm{STE}}}}^{\mathrm{col}}$ are the tuning parameters which control the contributions of the angle sparsity levels. It can be seen that (\ref{eqn:LS estimation l1 norm}) still remains challenging to solve due to the highy non-convex low-rank constraints and the coupled optimization variables. To tackle these difficulties, we propose a mixed channel MO algorithm.

\subsection{Mixed Channel MO Algorithm}\label{subsec:MV-MO Algorithm}
In problem (\ref{eqn:LS estimation l1 norm}), coupling exists not only between $\widehat{\mathbf{G}}_\mathrm{DTE}$ and $\widehat{\mathbf{H}}_{\mathrm{DTE},k}$ within the DTE-based channel but also between the DTE- and STE-based channels. To address this coupling among multiple variables in the mixed DTE-STE channels, we first decouple the DTE- and STE-based channels and optimize them alternately, a process referred to as the outer iteration. Specifically, during the outer iteration, we optimize $\widehat{\mathbf{H}}_{\mathrm{eq},k}^{\mathrm{STE}}$ in the STE-based channel while keeping the DTE-based channel fixed. Subsequently, when optimizing the DTE-based channel with the STE-based channel fixed, we further decouple the two variables within the DTE-based channel, i.e., $\widehat{\mathbf{G}}_\mathrm{DTE}$ and $\widehat{\mathbf{H}}_{\mathrm{DTE},k}$, and optimize them alternately. This process is referred to as the inner iteration. 

To effectively handle the low-rank constraints associated with these variables, we apply the MO method \cite{2022_Lin_channel_estimation}\cite{2022_Liu_ICC_2022}. Taking $\widehat{\mathbf{G}}_\mathrm{DTE}$ as an example, the feasible set can be represented as a typical Riemannian manifold defined as $\mathcal{M}_P \triangleq \left\{ \widehat{\mathbf{G}}_\mathrm{DTE} \in \mathbb{C}^{N_\mathrm{BS} \times N_\mathrm{DTE}} : \operatorname{rank}( \widehat{\mathbf{G}}_\mathrm{DTE} ) = P \right\}$.
% \begin{equation}\label{eqn:manifold}
%     \mathcal{M}_P \triangleq \left\{ \widehat{\mathbf{G}}_\mathrm{DTE} \in \mathbb{C}^{N_\mathrm{BS} \times N_\mathrm{DTE}} : \operatorname{rank}( \widehat{\mathbf{G}}_\mathrm{DTE} ) = P \right\}.
% \end{equation}
The crucial step in the MO method is the derivation of the Riemannian gradient, which can be deduced from the classic conjugate gradient in Euclidean space. By utilizing the property of matrix calculus as $\mathrm{d}(f)=\mathrm{Tr}\big(\nabla_{\mathbf{Z}^*}f\mathrm{d}(\mathbf{Z}^H)\big)$, 
% $\mathrm{d}(\|\mathbf{c}-\mathbf{AZb}\|^2)=\mathrm{Tr}\Big(\big(-\mathbf{A}^H\mathbf{c}\mathbf{b}^H+\mathbf{A}^H\mathbf{A}\mathbf{Z}\mathbf{b}\mathbf{b}^H\big)\mathrm{d}(\mathbf{Z}^H)\Big)$ and $\mathrm{d}(\|\ \boldsymbol{\lambda}_{\widehat{\mathbf{G}}}\|_1)=\frac{1}{2}\mathrm{Tr}\Big(\big(\mathbf{A}_{\mathrm{BS}} \mathbf{Y}_{\widehat{\mathbf{G}}}\mathbf{A}_{\mathrm{I}}^{H}\big)\mathrm{d}(\mathbf{X}^H)\Big)$,
for $\widehat{\mathbf{G}}_\mathrm{DTE}$, $\widehat{\mathbf{H}}_{\mathrm{DTE},k}$, and $\widehat{\mathbf{H}}_{\mathrm{eq},k}^{\mathrm{STE}}$, the Euclidean conjugate gradient of objective function in (\ref{eqn:LS estimation l1 norm}) can be expressed as
\begin{equation}\label{eqn:Euclidean conjugate gradient G_DTE}
\begin{array}{rl}
\hspace{0.45cm}\nabla_{{{\widehat{\mathbf{G}}_\mathrm{DTE}}^*}}f_{\widehat{\mathbf{G}}_\mathrm{DTE}}=&\sum\limits_{t=1}^{{T}}\Big(\big( \widehat{\mathbf{H}}_{\mathrm{eq},k}^{\mathrm{STE}}\mathbf{s}_{k}[t] - \mathbf{r}_{k}[t] \big)\mathbf{s}_{k}^H[t]\widehat{\mathbf{H}}_{\mathrm{DTE},k}^H\boldsymbol{\Phi}_{k}^H[t] + {\widehat{\mathbf{G}}_\mathrm{DTE}}\boldsymbol{\Phi}_{k}[t]\\
&\times \widehat{\mathbf{H}}_{\mathrm{DTE},k}\mathbf{s}_{k}[t]\mathbf{s}_{k}^H[t] \widehat{\mathbf{H}}_{\mathrm{DTE},k}^H\boldsymbol{\Phi}_{k}^H[t]\Big)+\frac{\upsilon_{\widehat{\mathbf{G}}_\mathrm{DTE}}}{2} \mathbf{A}_{\mathrm{BS}} \mathbf{Y}_{{\widehat{\mathbf{G}}_\mathrm{DTE}}} \mathbf{A}_{\mathrm{DTE}}^{H},
\end{array}
\end{equation}
\begin{equation}\label{eqn:Euclidean conjugate gradient H_DTE}
\begin{array}{rl}
\nabla_{{{\widehat{\mathbf{H}}_{\mathrm{DTE},k}}^*}}f_{{{\widehat{\mathbf{H}}_{\mathrm{DTE},k}}}}=&\sum\limits_{t=1}^{{T}}\Big( \boldsymbol{\Phi}_{k}^H[t]\widehat{\mathbf{G}}_\mathrm{DTE}^H \big( \widehat{\mathbf{H}}_{\mathrm{eq},k}^{\mathrm{STE}}\mathbf{s}_{k}[t]- \mathbf{r}_{k}[t] \big)\mathbf{s}_{k}^H[t]+ \boldsymbol{\Phi}_{k}^H[t]\widehat{\mathbf{G}}_\mathrm{DTE}^H\\
&\times {\widehat{\mathbf{G}}_\mathrm{DTE}}\boldsymbol{\Phi}_{k}[t]\widehat{\mathbf{H}}_{\mathrm{DTE},k}\mathbf{s}_{k}[t]\mathbf{s}_{k}^H[t] \Big)+\frac{\upsilon_{\widehat{\mathbf{H}}_{\mathrm{DTE},k}}}{2} \mathbf{A}_{\mathrm{DTE}} \mathbf{Y}_{\widehat{\mathbf{H}}_{\mathrm{DTE},k}} \mathbf{A}_{\mathrm{UE}}^{H},
\end{array}
\end{equation}
\begin{equation}\label{eqn:Euclidean conjugate gradient H_STE}
\begin{array}{rl}
\nabla_{\big(\widehat{\mathbf{H}}_{\mathrm{eq},k}^{\mathrm{STE}}\big)^*}f_{\widehat{\mathbf{H}}_{\mathrm{eq},k}^{\mathrm{STE}}}=&\sum\limits_{t=1}^{{T}}\Big( \big( \widehat{\mathbf{G}}_\mathrm{DTE}\boldsymbol{\Phi}_{k}[t]\widehat{\mathbf{H}}_{\mathrm{DTE},k}\mathbf{s}_{k}[t] - \mathbf{r}_{k}[t] \big)\mathbf{s}_{k}^H[t]+ \widehat{\mathbf{H}}_{\mathrm{eq},k}^{\mathrm{STE}}\mathbf{s}_{k}[t] \mathbf{s}_{k}^H[t]\Big)\\
& +\frac{\upsilon_{\widehat{\mathbf{H}}_{\mathrm{eq},k}^{\mathrm{STE}}}^{\mathrm{row}}}{2} \mathbf{A}_{\mathrm{BS}} \mathbf{Y}_{\widehat{\mathbf{H}}_{\mathrm{eq},k}^{\mathrm{STE}}}^{\mathrm{row}} \big(\mathbf{A}_{\mathrm{UE}} \mathbf{1}_{N_\mathrm{UE}}\big)^H+\frac{\upsilon_{\widehat{\mathbf{H}}_{\mathrm{eq},k}^{\mathrm{STE}}}^{\mathrm{col}}}{2} \big(\mathbf{A}_{\mathrm{BS}} \mathbf{1}_{N_\mathrm{BS}}\big)\mathbf{Y}_{\widehat{\mathbf{H}}_{\mathrm{eq},k}^{\mathrm{STE}}}^{\mathrm{col}} \mathbf{A}_{\mathrm{UE}}^H,
\end{array}
\end{equation}
where $\big[\mathbf{Y}_{\widehat{\mathbf{G}}_\mathrm{DTE}}\big]_{ij}=\big[\mathbf{A}_{\mathrm{BS}}^H{\widehat{\mathbf{G}}_\mathrm{DTE}}\mathbf{A}_{\mathrm{DTE}}\big]_{ij}\big/\big|\big[\mathbf{A}_{\mathrm{BS}}^H{\widehat{\mathbf{G}}_\mathrm{DTE}}\mathbf{A}_{\mathrm{DTE}}\big]_{ij}\big|$, $\big[\mathbf{Y}_{\widehat{\mathbf{H}}_{\mathrm{DTE},k}}\big]_{ij}=\big[\mathbf{A}_{\mathrm{DTE}}^H{\widehat{\mathbf{H}}_{\mathrm{DTE},k}}\mathbf{A}_{\mathrm{UE}}\big]_{ij}\big/$ $\big|\big[\mathbf{A}_{\mathrm{DTE}}^H{\widehat{\mathbf{H}}_{\mathrm{DTE},k}}\mathbf{A}_{\mathrm{UE}}\big]_{ij}\big|$, $\big[\mathbf{Y}_{\widehat{\mathbf{H}}_{\mathrm{eq},k}^{\mathrm{STE}}}^{\mathrm{row}}\big]_{i}=\big[\mathbf{A}_{\mathrm{BS}}^H{\widehat{\mathbf{H}}_{\mathrm{eq},k}^{\mathrm{STE}}}\big(\mathbf{A}_{\mathrm{UE}}\mathbf{1}_{\mathrm{UE}}\big)\big]_{i}\big/\big|\big[\mathbf{A}_{\mathrm{BS}}^H{\widehat{\mathbf{H}}_{\mathrm{eq},k}^{\mathrm{STE}}}\big(\mathbf{A}_{\mathrm{UE}}\mathbf{1}_{\mathrm{UE}}\big)\big]_{i}\big|$, $\big[\mathbf{Y}_{\widehat{\mathbf{H}}_{\mathrm{eq},k}^{\mathrm{STE}}}^{\mathrm{col}}\big]_{j}=$ $ \big[\big($ $\mathbf{1}_\mathrm{BS}^T\mathbf{A}_{\mathrm{BS}}^H\big){\widehat{\mathbf{H}}_{\mathrm{eq},k}^{\mathrm{STE}}}\mathbf{A}_{\mathrm{UE}}\big]_{j}\big/\big|\big[\big(\mathbf{1}_\mathrm{BS}^T\mathbf{A}_{\mathrm{BS}}^H\big){\widehat{\mathbf{H}}_{\mathrm{eq},k}^{\mathrm{STE}}}\mathbf{A}_{\mathrm{UE}}\big]_{j}\big|$. Based on (\ref{eqn:Euclidean conjugate gradient G_DTE}), (\ref{eqn:Euclidean conjugate gradient H_DTE}), and (\ref{eqn:Euclidean conjugate gradient H_STE}), we can employ the MO method to estimate $\widehat{\mathbf{H}}_{\mathrm{eq},k}^{\mathrm{STE}}$, $\widehat{\mathbf{G}}_\mathrm{DTE}$, and $\widehat{\mathbf{H}}_{\mathrm{DTE},k}$ with the two levels of iterations mentioned above. The proposed mixed channel MO algorithm is summarized in \textbf{Algorithm 1}.

With the DSD channel model and the mixed channel MO algorithm, we can obtain the CSI for all UEs in the HE-IRS-assisted system, and thus complete the DSD-MO estimation scheme.
\begin{algorithm}[t]
	\caption{Mixed channel MO algorithm}
\hspace*{\algorithmicindent} {\color{black}\textbf{Input:} $\upsilon_{\widehat{\mathbf{G}}_\mathrm{DTE}}$, $\upsilon_{\widehat{\mathbf{H}}_{\mathrm{DTE},k}}$, $\upsilon_{\widehat{\mathbf{H}}_{\mathrm{eq},k}^{\mathrm{STE}}}^{\mathrm{row}}$, $\upsilon_{{\widehat{\mathbf{H}}_{\mathrm{eq},k}^{\mathrm{STE}}}}^{\mathrm{col}}$, $P$, $Q$, $\mu$, $\mathbf{s}_{k}[t]$, $\mathbf{r}_{k}[t]$, $\mathbf{\Phi}_{k}[t]$.}
	\begin{algorithmic}[1]
	\STATE Randomly initialize ${\widehat{\mathbf{G}}_\mathrm{DTE}}^{(0)}$, $\widehat{\mathbf{H}}_{\mathrm{DTE},k}^{(0,0)}$, and $\big({\widehat{\mathbf{H}}_{\mathrm{eq},k}^{\mathrm{STE}}}\big)^{(0,0)}$. Set $i=0$ and $j=0$.
\REPEAT
    \STATE $i\leftarrow i+1$.
    \STATE Update $\big({\widehat{\mathbf{H}}_{\mathrm{eq},k}^{\mathrm{STE}}}\big)^{(i)}$ for given ${\widehat{\mathbf{G}}_\mathrm{DTE}}^{(i-1,j)}$ and $\widehat{\mathbf{H}}_{\mathrm{DTE},k}^{(i-1,j)}$ using the MO method.
    \REPEAT
        \STATE $j\leftarrow j+1$.
        \STATE Update ${\widehat{\mathbf{G}}_\mathrm{DTE}}^{(i-1,j)}$ for given $\widehat{\mathbf{H}}_{\mathrm{DTE},k}^{(i-1,j-1)}$ and $\big({\widehat{\mathbf{H}}_{\mathrm{eq},k}^{\mathrm{STE}}}\big)^{(i)}$ using the MO method.
        \STATE Update $\widehat{\mathbf{H}}_{\mathrm{DTE},k}^{(i-1,j)}$ for given ${\widehat{\mathbf{G}}_\mathrm{DTE}}^{(i-1,j)}$ and $\big({\widehat{\mathbf{H}}_{\mathrm{eq},k}^{\mathrm{STE}}}\big)^{(i)}$ using the MO method.
    \UNTIL The inner stopping criterion is met.
\UNTIL The outer stopping criterion is met.
	\end{algorithmic}
 \hspace*{\algorithmicindent} {\color{black}\textbf{Output:} $\widehat{\mathbf{H}}_{\mathrm{ca},k}^{\mathrm{DTE}}=\widehat{\mathbf{H}}_{\mathrm{DTE},k}^T \odot \widehat{\mathbf{G}}_\mathrm{DTE}$, $\widehat{\mathbf{H}}_{\mathrm{eq},k}^{\mathrm{STE}}$.}
\end{algorithm}

\subsection{Robust Rank Selection Rule with Rank Mismatches}\label{subsec:Rank Selection Rule with Mismatched Path Numbers}
In the previous discussion, we assumed the perfect knowledge of the numbers of propagation paths in all channels. In practice, achieving such knowledge is infeasible. 
% Traditional direction of arrival (DoA) estimation methods, such as multiple signal classification (MUSIC) \cite{2015_Vallet_DoA_music}\cite{2024_Chen_DoA_music} and estimation of signal parameters via rotational invariance techniques (ESPRIT) \cite{2015_Wu_DoA_ESPRIT}\cite{2005_Han_DoA_ESPRIT}, can provide approximate estimates of these path numbers, but they do not guarantee perfect accuracy. 
Consequently, the estimated path numbers often differ from the actual ones, a scenario referred to as mismatched path numbers.
% Without loss of generality, we assume that the mismatched path numbers for the channels from the HE-IRS to the BS and from the $k$-th UE to the HE-IRS can be expressed as 
% \begin{equation} \label{eqn:mismatch P and Q}
%     \widehat{P} = P+ \Delta P, \quad \widehat{Q} = Q+ \Delta Q,
% \end{equation}
% where $\Delta P \in \big[ \Delta P_{\mathrm{min}}, \Delta P_{\mathrm{max}} \big]$ and $\Delta Q \in \big[ \Delta Q_{\mathrm{min}}, \Delta Q_{\mathrm{max}} \big]$ represent the estimation errors for the path numbers $P$ and $Q$, respectively. We assume that the ranges $\big[ \Delta P_{\mathrm{min}}, \Delta P_{\mathrm{max}} \big]$ and $\big[ \Delta Q_{\mathrm{min}}, \Delta Q_{\mathrm{max}} \big]$ are known based on DoA estimation methods or prior observations. 
When the path numbers are mismatched, the constraints in (\ref{eqn:LS estimation}) are affected, potentially resulting in rank and angle sparsity mismatches. For channel angle sparsity mismatches, the $\ell_{1}$-norm regularization allows these constraints to be incorporated into the objective function as regularization terms, which can be implicitly addressed, as shown in (\ref{eqn:LS estimation l1 norm}). 

However, for channel rank mismatches, the situation is different. In the MO-based method, the feasible solution space for each channel is defined by its fixed-rank manifold. Therefore, selecting an appropriate fixed rank for each channel is critical to ensuring estimation accuracy. The rule for selecting the fixed channel rank during the estimation process is established in the following two lemmas.
% For angle sparsity mismatch, by applying the $\ell_{1}$-norm regularization to relax the $\ell_{0}$-norm constraints associated with angle sparsity, these constraints are incorporated into the objective function as regularization terms, which allows the mismatched angular sparsity to be implicitly addressed, as shown in (\ref{eqn:LS estimation l1 norm}). 
% Thus, we will focus on the rank selection rule.

% For rank mismatch, it is not the case. In the MO-based method, the feasible solution space for each channel is defined by its fixed-rank manifold, as illustrated in (\ref{eqn:manifold}). Thus, selecting the appropriate fixed rank for each channel is crucial for ensuring the accuracy of its estimate. The selection rule of the fixed rank during the estimation process will be established in the following two Lemmas.
% the Since the rank of each channel is directly related to the number of its propagation paths, a mismatch in the path number will lead to a corresponding mismatch in rank. Moreover, in the MO-based method, the feasible solution space for each channel is defined by its fixed-rank manifold, as illustrated in (\ref{eqn:manifold}). Therefore, selecting the appropriate fixed rank for each channel is crucial for ensuring the accuracy of its estimate. The selection rule of the fixed rank during the estimation process will be established in the following two Lemmas.
\begin{lemma}\label{lem:low rank representation}
    Let $\mathbf{H} \in \mathbb{C}^{N_1 \times N_2}$ be a matrix with $\mathrm{rank}\big( \mathbf{H} \big) \le m$, where $m \le \mathrm{min}\big\{N_1,N_2\big\}$. Then, for any $\epsilon > 0$, there exists a matrix $\widetilde{\mathbf{H}} \in \mathbb{C}^{N_1 \times N_2}$ with $\mathrm{rank}\big( \widetilde{\mathbf{H}} \big) = m$ such that $\big\| \mathbf{H}-\widetilde{\mathbf{H}} \big\| < \epsilon $.
\end{lemma}
\emph{Proof}: Using singular value decomposition (SVD), the matrix $\mathbf{H}$ can be expressed as
$\mathbf{H}=\mathbf{U}_m\boldsymbol{\Sigma}_{\mathbf{H}}\mathbf{V}_m^H$, where $\mathbf{U}_m \in \mathbb{C}^{N_1 \times m}$ and $\mathbf{V}_m \in \mathbb{C}^{N_2 \times m}$ are constructed by extracting the first $m$ columns from the standard left and right singular vector matrices, respectively, and $\boldsymbol{\Sigma}_{\mathbf{H}} \in \mathbb{C}^{m \times m}$ is a diagonal matrix of singular values given by
\begin{equation} \label{low rank original diagnal matrix}
    \boldsymbol{\Sigma}_{\mathbf{H}}=\mathrm{diag}\big( \sigma_1,\sigma_2,\dots,\sigma_r,0,\dots,0 \big),
\end{equation}
where $\sigma_1\ge\sigma_2\ge\dots\ge\sigma_r>0$ and $r=\mathrm{rank}\big(\mathbf{H}\big)$. To construct $\widetilde{\mathbf{H}}$ with $\mathrm{rank}\big( \widetilde{\mathbf{H}} \big) = m$, we define a new diagonal matrix $\boldsymbol{\Sigma}_{\widetilde{\mathbf{H}}}\in \mathbb{C}^{m \times m}$ as
\begin{equation} \label{low rank construct diagnal matrix}
    \boldsymbol{\Sigma}_{\widetilde{\mathbf{H}}}=\mathrm{diag}\big( \sigma_1,\sigma_2,\dots,\sigma_r,\varepsilon_{1},\varepsilon_{2}\dots,\varepsilon_{m-r} \big),
\end{equation}
where the first $r$ singular values are identical to those of $\boldsymbol{\Sigma}_{\mathbf{H}}$, and the next $m-r$ singular values, $\varepsilon_{1},\varepsilon_{2},\dots,\varepsilon_{m-r}$, are arbitrarily small numbers satisfying $\varepsilon_{1}\ge\varepsilon_{2}\ge\dots\ge\varepsilon_{m-r}>0$. By utilizing this modified singular value matrix, we construct $\widetilde{\mathbf{H}}$ as $\widetilde{\mathbf{H}}=\mathbf{U}_m\boldsymbol{\Sigma}_{\widetilde{\mathbf{H}}}\mathbf{V}_m^H \in \mathbb{C}^{N_1 \times N_2}$. The difference between $\mathbf{H}$ and $\widetilde{\mathbf{H}}$ can then be calculated as
\begin{equation} \label{low rank approximate}
     \big\|\mathbf{H}-\widetilde{\mathbf{H}}\big\| = \big\| \mathbf{U}_m\big(\boldsymbol{\Sigma}_{\mathbf{H}}-\boldsymbol{\Sigma}_{\widetilde{\mathbf{H}}}\big)\mathbf{V}_m^H \big\| = \big\|\boldsymbol{\Sigma}_{\mathbf{H}}-\boldsymbol{\Sigma}_{\widetilde{\mathbf{H}}}\big\|,
\end{equation}
where (\ref{low rank approximate}) follows from the property  $\mathbf{U}_m^H\mathbf{U}_m=\mathbf{V}_m^H\mathbf{V}_m=\mathbf{I}_m$. Based on (\ref{low rank original diagnal matrix}) and (\ref{low rank construct diagnal matrix}), this difference can be further given by $\big\|\boldsymbol{\Sigma}_{\mathbf{H}}-\boldsymbol{\Sigma}_{\widetilde{\mathbf{H}}}\big\| = \sqrt{\sum\nolimits_{i=1}^{m-r}{\varepsilon_i^2}}$. Thus, for any $\epsilon > 0$, we can choose $\varepsilon_{1},\varepsilon_{2},\dots,\varepsilon_{m-r}$ such that $\sqrt{\sum\nolimits_{i=1}^{m-r}{\varepsilon_i^2}}<\epsilon$. This completes the proof of Lemma 2.$\hfill\blacksquare$

\begin{lemma}\label{lem:high rank representation}
    Let $\mathbf{H} \in \mathbb{C}^{N_1 \times N_2}$ be a matrix with $\mathrm{rank}\big( \mathbf{H} \big) > m$, where $m < \mathrm{min}\big\{N_1,N_2\big\}$. Then, for any matrix $\widetilde{\mathbf{H}} \in \mathbb{C}^{N_1 \times N_2}$ with $\mathrm{rank}\big( \widetilde{\mathbf{H}} \big) = m$, there exists an $\epsilon > 0$ such that $\big\| \mathbf{H}-\widetilde{\mathbf{H}} \big\| \ge \epsilon $.
\end{lemma}
\emph{Proof}: Using proof by contradiction, assume that for any matrix $\widetilde{\mathbf{H}} \in \mathbb{C}^{N_1 \times N_2}$ with $\mathrm{rank}\big( \widetilde{\mathbf{H}} \big) = m$, there does not exist an $\epsilon > 0$ such that $\big\| \mathbf{H}-\widetilde{\mathbf{H}} \big\| \ge \epsilon $. This implies $\big\| \mathbf{H}-\widetilde{\mathbf{H}} \big\| = 0$ for all such $\widetilde{\mathbf{H}}$. Consequently, we would have $\widetilde{\mathbf{H}}=\mathbf{H}$, leading to the conclusion $\mathrm{rank}\big( \widetilde{\mathbf{H}} \big)=\mathrm{rank}\big( \mathbf{H} \big)=m$. This contradicts the assumption $\mathrm{rank}\big( \mathbf{H}\big)>m$. Thus, the proof is complete.$\hfill\blacksquare$

From Lemmas 2 and 3, it can be observed that for a given fixed-rank matrix, there always exists a matrix with a higher rank that can approximate the original matrix arbitrarily closely. Conversely, no matrix with a lower rank can achieve such arbitrary approximation. This implies that in scenarios with channel rank mismatches, using an estimated rank that is lower than the true rank as a fixed-rank constraint may limit the ability to approximate the actual feasible solution space, thereby degrading the estimation performance. To address this issue, it is advisable to adopt a little higher rank, rather than the directly estimated one, as the fixed-rank constraint for MO method, which serves as the robust rank selection rule.

\section{Two-Stage Beamforming Optimization}\label{sec:Two-Stage Beamforming Optimization}
Given the fact that the HE-IRS consists of two types of reflecting elements with both static and dynamic phase-tuning capabilities, the beamforming optimization for the STEs and DTEs is divided into two stages. In this section, we first outline the strategy for the two-stage beamforming optimization, and then propose two algorithms for the two stages, respectively.
% In this section, we first explain the strategy of the two-stage beamforming optimization for HE-IRS. Subsequently, we present the proposed detailed algorithms in offline and online stage.
% In this section, we propose a beamforming strategy for the HE-IRS-assisted multi-user MIMO system. Specifically, in the offline stage, a WBS-MO algorithm targeting the whole UE location area is proposed to optimized the phase shifts of the STEs. In the online stage, with the estimated HE-IRS channel by the DSD-MO scheme and the optimized phase shifts , a WMMSE-EI algorithm targeting specific UEs is proposed to dynamically optimize beamforming design for the BS and the DTEs.
\subsection{Two-stage Beamforming Strategy}\label{subsec:The HE-IRS Beamforming Strategy}
% Given the HE-IRS consisting of two types of reflecting elements with both static and dynamic phase-shifting capabilities, the beamforming optimization for the STEs and DTEs is divided into the following two stages:
In the offline stage, the objective is to optimize the phase shifts of the STEs. Since their phase shifts cannot be dynamically adjusted, traditional phase optimization methods targeting specific UEs are not suitable. Inspired by \cite{2017_Song_wide_beam}\cite{2024_Al-Tous_SRS}, we propose an offline beamforming strategy that concentrates the reflected beam power on the entire UE location area.
% Instead, we adopt a strategy of \textit{de-focusing} the reflected beam power to achieve broad coverage of the UE location area with a wide beam \cite{2017_Song_wide_beam}\cite{2024_Al-Tous_SRS}. 
However, due to the Heisenberg uncertainty principle, a sequence cannot simultaneously be beam-limited (angular domain) and index-limited (spatial domain) \cite{1978_Slepian_heisenverg_uncertainty}. To concentrate power on the UE location area while adhering to the unit-modulus constraint imposed by the STE phase shifters, we propose a wide beam synthesis-MO (WBS-MO) algorithm.

In the online stage, with the pre-determined phase shifts of the STEs, we propose a weighted mean square error minimization-element iteration (WMMSE-EI) algorithm to optimize the beamforming for the BS and the DTEs based on the estimated HE-IRS channel using the DSD-MO scheme. This allows the beams generated by the DTEs to be dynamically coordinated with those produced by the STEs, resulting in synthesized directive beams for the specific UEs. 
% The two proposed algorithms, designed for the offline and online stages, will be presented in detail in the following subsections.

\subsection{Offline Beamforming: WBS-MO Algorithm}\label{subsec:Offline Stage: WBS-MO Algorithm}
Before deploying the HE-IRS, the phase shifts of the STEs are optimized by simulating the STE array at its intended deployment location to facilitate communication. For simplicity, it is assumed that communication primarily occurs via the LoS path between the BS and the STE array, as well as those between the STE array and the UEs. Specifically, since the STE array and the BS will not move once deployment, the azimuth and elevation AoAs of the HE-IRS, i.e., $\theta_\mathrm{r}^0$ and $\varphi_\mathrm{r}^0$, are considered fixed. Meanwhile, the UEs are distributed within an designated angular space defined as $\mathcal{A}=\big\{ \big(\vartheta_{\mathrm{t}}^0,\zeta_{\mathrm{t}}^0\big) \big| \space \vartheta_{\mathrm{t}}^0\in \big[\vartheta_{\mathrm{min}},  \vartheta_{\mathrm{max}} \big],\space \zeta_{\mathrm{t}}^0 \in \big[\zeta_{\mathrm{min}}, \zeta_{\mathrm{max}} \big]  \big\}$, where $\vartheta_{\mathrm{t}}^0$ and $\zeta_{\mathrm{t}}^0$ are the azimuth and elevation AoDs of the HE-IRS. Thus, for a direction vector $\boldsymbol{\theta}=\big( \theta_\mathrm{r}^0,\theta_\mathrm{r}^0,\vartheta_{\mathrm{t}}^0,\zeta_{\mathrm{t}}^0\big)$, the directivity gain of the STE array is given by
\begin{equation} \label{eqn:gain}
    G\big(\boldsymbol{\theta}, \boldsymbol{\Omega} \big) = \big| \mathbf{a}_\mathrm{STE}^H \big( \vartheta_{\mathrm{t}}^0, \zeta_{\mathrm{t}}^0 \big)\boldsymbol{\Omega} \mathbf{a}_\mathrm{STE}\big( \theta_\mathrm{r}^0, \varphi_\mathrm{r}^0 \big) \big|^2,
\end{equation}
where  $\mathbf{a}_\mathrm{STE}\big( \theta_\mathrm{r}^0, \varphi_\mathrm{r}^0 \big)=\mathbf{a}_y\big(\theta_\mathrm{r}^0, \varphi_\mathrm{r}^0\big) \otimes \mathbf{a}_z\big(\varphi_\mathrm{r}^0\big)$ and $\mathbf{a}_\mathrm{STE}\big( \vartheta_{\mathrm{t}}^0, \zeta_{\mathrm{t}}^0 \big)=\mathbf{a}_y\big(\vartheta_{\mathrm{t}}^0,\zeta_{\mathrm{t}}^0) \otimes \mathbf{a}_z(\zeta_{\mathrm{t}}^0)$. By maximizing the cumulative directivity gain over the angular space $\mathcal{A}$, the wide beam synthesis optimization problem is formulated as
\begin{equation}\label{eqn:WBS problem}
    \begin{array}{cl}
    \underset{\boldsymbol{\Omega}}{\operatorname{maximize}} & D_{\mathcal{A}}=\iint\limits_{\mathcal{A}}{G\big(\boldsymbol{\theta}, \boldsymbol{\Omega} \big)} d\zeta_{\mathrm{t}}^0 d\vartheta_{\mathrm{t}}^0 \\
    \text { subject to } & \big| \big[ \boldsymbol{\Omega} \big] _{nn} \big|=1,\space\forall n.
\end{array}
\end{equation}

Observing the objective function $D_{\mathcal{A}}$ in problem (\ref{eqn:WBS problem}), it is evident that the highly coupled angular parameters significantly complicate $G\big(\boldsymbol{\theta}, \boldsymbol{\Omega} \big)$, making it difficult to evaluate the double integral and obtain an analytical solution. To address this issue, we rewrite $G\big(\boldsymbol{\theta}, \boldsymbol{\Omega} \big)$ as 
\begin{equation} \label{eqn:G diag transform}
    G\big(\boldsymbol{\theta}, \boldsymbol{\Omega} \big) = \big| \boldsymbol{\omega}^H \cdot \widetilde{\mathbf{a}}_\mathrm{STE}\big(\boldsymbol{\theta}\big)\big|^2,
\end{equation}
where $\widetilde{\mathbf{a}}_\mathrm{STE}\big(\boldsymbol{\theta}\big)=\mathbf{a}_\mathrm{STE}^{*}\big( \theta_\mathrm{r}^0, \varphi_\mathrm{r}^0 \big) \circ \mathbf{a}_\mathrm{STE}\big( \vartheta_{\mathrm{t}}^0, \zeta_{\mathrm{t}}^0 \big)$. The term $\widetilde{\mathbf{a}}_\mathrm{STE}\big(\boldsymbol{\theta}\big)$ is then further decomposed along the y- and z-directions as 
\begin{equation} \label{eqn:a_STE yz}
\begin{array}{rl}
     \widetilde{\mathbf{a}}_\mathrm{STE}\big(\boldsymbol{\theta}\big) \overset{\phantom{(a)}}{=} & \big[ \mathbf{a}_y^*\big(\theta_\mathrm{r}^0, \varphi_\mathrm{r}^0\big) \otimes \mathbf{a}_z^*\big(\varphi_\mathrm{r}^0\big) \big] \circ \big[ \mathbf{a}_y\big(\vartheta_{\mathrm{t}}^0,\zeta_{\mathrm{t}}^0 \big) \otimes \mathbf{a}_z \big(\zeta_{\mathrm{t}}^0 \big) \big] \\
     \overset{(a)}{=} & \big[ \mathbf{a}_y^*\big(\theta_\mathrm{r}^0, \varphi_\mathrm{r}^0\big) \odot \mathbf{a}_z^*\big(\varphi_\mathrm{r}^0\big) \big] \circ \big[ \mathbf{a}_y\big(\vartheta_{\mathrm{t}}^0,\zeta_{\mathrm{t}}^0 \big) \odot \mathbf{a}_z \big(\zeta_{\mathrm{t}}^0 \big) \big] \\
     \overset{(b)}{=} & \big[ \mathbf{a}_y^*\big(\theta_\mathrm{r}^0, \varphi_\mathrm{r}^0\big) \circ \mathbf{a}_y\big(\vartheta_{\mathrm{t}}^0,\zeta_{\mathrm{t}}^0 \big) \big] \odot \big[ \mathbf{a}_z^*\big(\varphi_\mathrm{r}^0\big) \circ \mathbf{a}_z \big(\zeta_{\mathrm{t}}^0 \big) \big] \\
     \overset{(c)}{=} & \mathbf{b}_y\big(\rho_\mathrm{y}\big) \otimes \mathbf{b}_z\big(\rho_\mathrm{z}\big),
\end{array}
\end{equation}
where $\mathbf{b}_y\big(\rho_\mathrm{y}\big)=\mathbf{a}_y^*\big(\theta_\mathrm{r}^0, \varphi_\mathrm{r}^0\big) \circ \mathbf{a}_y\big(\vartheta_{\mathrm{t}}^0,\zeta_{\mathrm{t}}^0 \big)$ and $\mathbf{b}_z\big(\rho_\mathrm{z}\big)=\mathbf{a}_z^*\big(\varphi_\mathrm{r}^0\big) \circ \mathbf{a}_z \big(\zeta_{\mathrm{t}}^0 \big)$. Here, $\rho_\mathrm{y}=\sin\big(\vartheta_{\mathrm{t}}^0\big)\sin\big(\zeta_{\mathrm{t}}^0\big)-\sin\big(\theta_\mathrm{r}^0\big)\sin\big(\varphi_\mathrm{r}^0\big)\in \big[\rho_\mathrm{y}^\mathrm{min}, \rho_\mathrm{y}^\mathrm{max} \big]$, $\rho_\mathrm{z}=\cos\big(\zeta_{\mathrm{t}}^0\big)-\cos\big(\varphi_\mathrm{r}^0\big)\in \big[\rho_\mathrm{z}^\mathrm{min}, \rho_\mathrm{z}^\mathrm{max} \big]$. (a) utilizes the equivalence $\mathbf{a}\otimes\mathbf{b}=\mathbf{a}\odot\mathbf{b}$ for column vectors $\mathbf{a}$ and $\mathbf{b}$. (b) follows from the relationship between the Khatri-Rao and Hadamard products, which states that $\big(\mathbf{A} \odot \mathbf{B}\big)\circ\big(\mathbf{C} \odot \mathbf{D}\big)=\big(\mathbf{A}\circ\mathbf{C}\big)\odot\big(\mathbf{B}\circ\mathbf{D}\big)$, with arbitrary matrices $\mathbf{A}$, $\mathbf{B}$, $\mathbf{C}$, and $\mathbf{D}$. (c) results from the same equivalence as (a). Under the assumption of the UPA STEs, we further express $\boldsymbol{\omega}=\boldsymbol{\omega}_\mathrm{y} \otimes \boldsymbol{\omega}_\mathrm{z}$, with $\boldsymbol{\omega}_\mathrm{y}\in\mathbb{C}^{N_{\mathrm{STE}}^\mathrm{y}}$ and $\boldsymbol{\omega}_\mathrm{z}\in\mathbb{C}^{N_\mathrm{z}}$ denoting the phase shifters in the STE array along the y- and z- directions, respectively. Substituting $\widetilde{\mathbf{a}}_\mathrm{STE}\big(\boldsymbol{\theta}\big)=\mathbf{b}_y\big(\rho_\mathrm{y}\big) \otimes \mathbf{b}_z\big(\rho_\mathrm{z}\big)$ and $\boldsymbol{\omega}=\boldsymbol{\omega}_\mathrm{y} \otimes \boldsymbol{\omega}_\mathrm{z}$ into (\ref{eqn:G diag transform}), we obtain
\begin{equation} \label{eqn:KR Har}
\begin{array}{rl}
     G\big(\boldsymbol{\theta}, \boldsymbol{\Omega} \big) \overset{\phantom{(a)}}{=} & \big| \big(\boldsymbol{\omega}_\mathrm{y}^H \otimes \boldsymbol{\omega}_\mathrm{z}^H \big)  \big(\mathbf{b}_y\big(\rho_\mathrm{y}\big) \otimes \mathbf{b}_z\big(\rho_\mathrm{z}\big)\big) \big|^2 \\
      \overset{(d)}{=} & \big(\boldsymbol{\omega}_\mathrm{y}^H \mathbf{b}_y\big(\rho_\mathrm{y}\big) \mathbf{b}_y^H\big(\rho_\mathrm{y}\big)\boldsymbol{\omega}_\mathrm{y}\big) \big(\boldsymbol{\omega}_\mathrm{z}^H \mathbf{b}_z\big(\rho_\mathrm{z}\big) \mathbf{b}_z^H\big(\rho_\mathrm{z}\big)\boldsymbol{\omega}_\mathrm{z}\big) \\
      \overset{\phantom{(e)}}{=} & \big(\boldsymbol{\omega}_\mathrm{y}^H \mathbf{B}_\mathrm{y}\big(\rho_\mathrm{y}\big) \boldsymbol{\omega}_\mathrm{y}  \big)\big(\boldsymbol{\omega}_\mathrm{z}^H \mathbf{B}_\mathrm{z}\big( \rho_\mathrm{z}\big)\boldsymbol{\omega}_\mathrm{z}  \big),
\end{array}
\end{equation}
where $\mathbf{B}_\mathrm{y}\big(\rho_\mathrm{y}\big)=\mathbf{b}_\mathrm{y}^H\big(\rho_\mathrm{y}\big)\mathbf{b}_\mathrm{y}\big(\rho_\mathrm{y}\big)$ and $\mathbf{B}_\mathrm{z}\big( \rho_\mathrm{z}\big)=\mathbf{b}_\mathrm{z}^H\big( \rho_\mathrm{z}\big)\mathbf{b}_\mathrm{z}\big( \rho_\mathrm{z}\big)$. (d) follows from the property $\big(\mathbf{A}\otimes\mathbf{C}\big)\big(\mathbf{B}\otimes\mathbf{D}\big)=\big(\mathbf{A}\mathbf{B}\big)\otimes\big(\mathbf{C}\mathbf{D}\big)$ with arbitrary matrices $\mathbf{A}$, $\mathbf{B}$, $\mathbf{C}$, and $\mathbf{D}$.

As observed in (\ref{eqn:KR Har}), $G\big(\boldsymbol{\theta}, \boldsymbol{\Omega} \big)$ is decomposed into y- and z-direction gains, i.e., $\boldsymbol{\omega}_\mathrm{y}^H \mathbf{B}_\mathrm{y}\big(\rho_\mathrm{y}\big) \boldsymbol{\omega}_\mathrm{y}$ and $\boldsymbol{\omega}_\mathrm{z}^H \mathbf{B}_\mathrm{z}\big( \rho_\mathrm{z}\big)\boldsymbol{\omega}_\mathrm{z}$, respectively. These gains correspond to their angular parameters, $\rho_\mathrm{y}$ and $\rho_\mathrm{z}$. By defining a new angular space as $\mathcal{B}=\big\{ \big(\rho_\mathrm{y},\rho_\mathrm{z}\big) \big| \space \rho_\mathrm{y}\in \big[\rho_\mathrm{y}^\mathrm{min}, \rho_\mathrm{y}^\mathrm{max} \big],\space \rho_\mathrm{z} \in \big[\rho_\mathrm{z}^\mathrm{min}, \rho_\mathrm{z}^\mathrm{max} \big] \big\}$, the cumulative directivity gain across $\mathcal{B}$ is expressed as
\begin{equation}\label{eqn:gain upper bound}
    \begin{array}{rl}
    D_\mathcal{B} = & \iint\limits_{\mathcal{B}}{\big(\boldsymbol{\omega}_\mathrm{y}^H \mathbf{B}_\mathrm{y}\big(\rho_\mathrm{y}\big) \boldsymbol{\omega}_\mathrm{y}  \big)\big(\boldsymbol{\omega}_\mathrm{z}^H \mathbf{B}_\mathrm{z}\big( \rho_\mathrm{z}\big)\boldsymbol{\omega}_\mathrm{z}  \big)} d\rho_\mathrm{y} d\rho_\mathrm{z} \\
    \overset{(e)}{\le} & \boldsymbol{\omega}_\mathrm{y}^H \left[\int_{\rho_\mathrm{y}=\rho_\mathrm{y}^{\mathrm{min}}}^{\rho_\mathrm{y}^{\mathrm{max}}} {\mathbf{B}_\mathrm{y}\big(\rho_\mathrm{y}\big) d\rho_\mathrm{y}} \right] \boldsymbol{\omega}_\mathrm{y} + \boldsymbol{\omega}_\mathrm{z}^H \left[\int_{\rho_\mathrm{z}=\rho_\mathrm{z}^{\mathrm{min}}}^{\rho_\mathrm{z}^{\mathrm{max}}} {\mathbf{B}_\mathrm{z}\big(\rho_\mathrm{z}\big) d\rho_\mathrm{z}} \right] \boldsymbol{\omega}_\mathrm{z} \\
    \overset{\phantom{(f)}}{=} & \boldsymbol{\omega}_\mathrm{y}^H \boldsymbol{\Xi
    }_\mathrm{y} \boldsymbol{\omega}_\mathrm{y} + \boldsymbol{\omega}_\mathrm{z}^H \boldsymbol{\Xi}_\mathrm{z} \boldsymbol{\omega}_\mathrm{z},
\end{array}
\end{equation}
where $\boldsymbol{\Xi}_\mathrm{y}$ is expressed in (\ref{integraing y}), and the expression for $\boldsymbol{\Xi}_\mathrm{z}$ is analogous to that of  $\boldsymbol{\Xi}_\mathrm{y}$. (e) is derived using the Cauchy-Schwarz inequality. According to (\ref{eqn:gain upper bound}), the cumulative directivity gain in the $y$- and $z$-directions, $\boldsymbol{\omega}_\mathrm{y}^H \boldsymbol{\Xi}_\mathrm{y} \boldsymbol{\omega}_\mathrm{y}$ and $\boldsymbol{\omega}_\mathrm{z}^H \boldsymbol{\Xi}_\mathrm{z} \boldsymbol{\omega}_\mathrm{z}$, are decoupled. Consequently, an efficient way is to optimize them separately. Taking optimizing $\boldsymbol{\omega}_\mathrm{y}$ as an example, its optimization problem can be formulated as
\begin{figure*}[!b] % [!b] 强制放在当前页面底部
\hrulefill % 添加一条横线（可选）
\begin{align}
    \boldsymbol{\Xi}_\mathrm{y} = & N_\mathrm{STE}^\mathrm{y} \begin{bmatrix}
        \rho_\mathrm{y}^{\mathrm{max}} - \rho_\mathrm{y}^{\mathrm{min}} & \frac{e^{-\mathrm{j}\pi\rho_\mathrm{y}^{\mathrm{max}}}-e^{-\mathrm{j}\pi\rho_\mathrm{y}^{\mathrm{min}}}}{-\mathrm{j}\pi} & \cdots & \frac{e^{-\mathrm{j}\pi (N_\mathrm{STE}^\mathrm{y}-1)\rho_\mathrm{y}^{\mathrm{max}}}-e^{-\mathrm{j}\pi (N_\mathrm{STE}^\mathrm{y}-1)\rho_\mathrm{y}^{\mathrm{min}}}}{-\mathrm{j}\pi(N_\mathrm{STE}^\mathrm{y}-1)} \\
        \frac{e^{\mathrm{j}\pi\rho_\mathrm{y}^{\mathrm{max}}}-e^{\mathrm{j}\pi\rho_\mathrm{y}^{\mathrm{min}}}}{\mathrm{j}\pi} & \rho_\mathrm{y}^{\mathrm{max}} - \rho_\mathrm{y}^{\mathrm{min}} & \cdots & \frac{e^{-\mathrm{j}\pi (N_\mathrm{STE}^\mathrm{y}-2)\rho_\mathrm{y}^{\mathrm{max}}}-e^{-\mathrm{j}\pi (N_\mathrm{STE}^\mathrm{y}-2)\rho_\mathrm{y}^{\mathrm{min}}}}{-\mathrm{j}\pi(N_\mathrm{STE}^\mathrm{y}-2)} \\
        \vdots & \vdots & \vdots & \vdots \\
        \frac{e^{\mathrm{j}\pi (N_\mathrm{STE}^\mathrm{y}-1)\rho_\mathrm{y}^{\mathrm{max}}}-e^{\mathrm{j}\pi (N_\mathrm{STE}^\mathrm{y}-1)\rho_\mathrm{y}^{\mathrm{min}}}}{\mathrm{j}\pi(N_\mathrm{STE}^\mathrm{y}-1)} & \cdots & \cdots & \rho_\mathrm{y}^{\mathrm{max}} - \rho_\mathrm{y}^{\mathrm{min}}
    \end{bmatrix}.  \label{integraing y}
    % \boldsymbol{\Xi}_\mathrm{z} = & N_\mathrm{STE}^\mathrm{z} \begin{bmatrix}
    %     \rho_\mathrm{z}^{\mathrm{max}} - \rho_\mathrm{z}^{\mathrm{min}} & \frac{e^{-\mathrm{j}\pi\rho_\mathrm{z}^{\mathrm{max}}}-e^{-\mathrm{j}\pi\rho_\mathrm{z}^{\mathrm{min}}}}{-\mathrm{j}\pi} & \cdots & \frac{e^{-\mathrm{j}\pi (N_\mathrm{STE}^\mathrm{z}-1)\rho_\mathrm{z}^{\mathrm{max}}}-e^{-\mathrm{j}\pi (N_\mathrm{STE}^\mathrm{z}-1)\rho_\mathrm{z}^{\mathrm{min}}}}{-\mathrm{j}\pi(N_\mathrm{STE}^\mathrm{z}-1)} \\
    %     \frac{e^{\mathrm{j}\pi\rho_\mathrm{z}^{\mathrm{max}}}-e^{\mathrm{j}\pi\rho_\mathrm{z}^{\mathrm{min}}}}{\mathrm{j}\pi} & \rho_\mathrm{z}^{\mathrm{max}} - \rho_\mathrm{z}^{\mathrm{min}} & \cdots & \frac{e^{-\mathrm{j}\pi (N_\mathrm{STE}^\mathrm{z}-2)\rho_\mathrm{z}^{\mathrm{max}}}-e^{-\mathrm{j}\pi (N_\mathrm{STE}^\mathrm{z}-2)\rho_\mathrm{z}^{\mathrm{min}}}}{-\mathrm{j}\pi(N_\mathrm{STE}^\mathrm{z}-2)} \\
    %     \vdots & \vdots & \vdots & \vdots \\
    %     \frac{e^{\mathrm{j}\pi (N_\mathrm{STE}^\mathrm{z}-1)\rho_\mathrm{z}^{\mathrm{max}}}-e^{\mathrm{j}\pi (N_\mathrm{STE}^\mathrm{z}-1)\rho_\mathrm{z}^{\mathrm{min}}}}{\mathrm{j}\pi(N_\mathrm{STE}^\mathrm{z}-1)} & \cdots & \cdots & \rho_\mathrm{z}^{\mathrm{max}} - \rho_\mathrm{z}^{\mathrm{min}}
    % \end{bmatrix} \label{integraing z}
\end{align}
\end{figure*}
\begin{equation}\label{eqn:WBS problem y}
    \begin{array}{cl}
    \underset{\boldsymbol{\omega}_\mathrm{y}}{\operatorname{minimize}} & - \ \boldsymbol{\omega}_\mathrm{y}^H \boldsymbol{\Xi
    }_\mathrm{y} \boldsymbol{\omega}_\mathrm{y} \\
    \text { subject to } & \big| \big[ \boldsymbol{\omega}_\mathrm{y} \big] _{n_1} \big|=1,\space\forall n_1.
\end{array}
\end{equation}
The feasible set of $\boldsymbol{\omega}_\mathrm{y}$ also lies on a well-known complex circle Riemanian manifold \cite{2016_Yu_MO}, i.e., $\mathcal{M} \triangleq \big\{ \boldsymbol{\omega}_\mathrm{y} \in \mathbb{C}^{N_\mathrm{STE}^\mathrm{y} \times 1} : \big| \big[ \boldsymbol{\omega}_\mathrm{y} \big] _{n_1} \big|=1, \forall n_1 \big\}$. The Euclidean conjugate gradient of the objective function in (\ref{eqn:WBS problem y}) is expressed as $\nabla_{\boldsymbol{\omega}_\mathrm{y}}f_{\boldsymbol{\omega}_\mathrm{y}}= -\boldsymbol{\Xi
}_\mathrm{y} \boldsymbol{\omega}_\mathrm{y}$. Using this derived Euclidean conjugate gradient, the MO method can be straightforwardly applied to optimize $\boldsymbol{\omega}_\mathrm{y}$ under the constant modulus constraint. Similarly, $\boldsymbol{\omega}_\mathrm{z}$ can be also optimized. Once both $\boldsymbol{\omega}_\mathrm{y}$ and $\boldsymbol{\omega}_\mathrm{z}$ are optimized, the phase shifts of the STEs are calculated as $\boldsymbol{\omega} = \boldsymbol{\omega}_\mathrm{y} \otimes \boldsymbol{\omega}_\mathrm{z}$, thereby completing the optimization process for the offline stage. 
% The total complexity is $\mathcal{O}(\xi_1(N_\mathrm{STE}^\mathrm{y})^2+\xi_2(N_\mathrm{z})^2+N_\mathrm{STE})$, where $\xi_1$ and $\xi_2$ are the iteration numbers for updating $\boldsymbol{\omega}_\mathrm{y}$ and $\boldsymbol{\omega}_\mathrm{z}$ during the MO method. As observed, if the STE array is assumed to be square, the complexity of the WBS-MO algorithm is proportional to $N_\mathrm{STE}$, making it a low-complexity solution.
% By decomposing the STE array and the cumulative directional gain into different directions, the WBS-MO algorithm provides a low-complexity solution. The detailed complexity analysis of this algorithm is presented in Section \ref{subsec:complexity}.

\subsection{Online Beamforming: WMMSE-EI Algorithm}\label{subsec:Online Stage: WMMSE-EI Algorithm}
After deploying the HE-IRS, as the phase shifts of the STEs are already determined, we need to optimize the beamforming of the BS and the DTEs. By using the DSD-MO estimation scheme proposed in Section \ref{sec:Channel Estimation Scheme},  $\mathbf{H}_{\mathrm{ca},k}^\mathrm{DTE}$ and ${\mathbf{H}_{\mathrm{eq},k}^\mathrm{STE}}$ are estimated. Then, the received signal at the $k$-th UE is expressed as 
\begin{equation} \label{eqn:downlink y}
    \mathbf{y}_k=\left(\mathrm{mat}\big(\mathbf{H}_{\mathrm{ca},k}^\mathrm{DTE} \boldsymbol{\phi}_\mathrm{d}^{*}\big) + {\mathbf{H}_{\mathrm{eq},k}^\mathrm{STE}}\right)^H\mathbf{V}\mathbf{s}_\mathrm{d} + \mathbf{n}_k,
\end{equation}
% \begin{equation} \label{eqn:downlink y}
%     \mathbf{y}_k=\left(\mathbf{H}_{\mathrm{DTE},k}^H \boldsymbol{\Phi}_{\mathrm{d}} \mathbf{G}_\mathrm{DTE}^H + \big({\mathbf{H}_{\mathrm{eq},k}^\mathrm{STE}}\big)^H\right)\mathbf{V}\mathbf{s}_\mathrm{d} + \mathbf{n}_k,
% \end{equation}
where $\mathrm{mat}(\cdot)$ denotes the operation that reshapes an $N_\mathrm{BS}N_\mathrm{UE}\times1$ vector to an $N_\mathrm{BS}\times N_\mathrm{UE}$ matrix. $\boldsymbol{\Phi}_{\mathrm{d}}=\mathrm{diag}(\boldsymbol{\phi}_{\mathrm{d}})\in\mathbb{C}^{N_\mathrm{DTE}\times N_\mathrm{DTE}}$ is the downlink phase shift matrix of the DTEs. $\mathbf{H}_{\mathrm{e},k}=\big(\mathrm{mat}\big(\mathbf{H}_{\mathrm{ca},k}^\mathrm{DTE} \boldsymbol{\phi}_\mathrm{d}^{*}\big) + {\mathbf{H}_{\mathrm{eq},k}^\mathrm{STE}}\big)^H\in\mathbb{C}^{N_{\mathrm{UE}}\times N_{\mathrm{BS}}}$ is the equivalent channel from the BS to the $k$-th UE.
$\mathbf{V} = [\mathbf{V}_1,\dots,\mathbf{V}_K]\in\mathbb{C}^{N_{\mathrm{BS}}\times N_{\mathrm{s}}K}$ is the BS precoder, $\mathbf{s}_{\mathrm{d}}=[\mathbf{s}_{\mathrm{d},1}^H,\dots,\mathbf{s}_{\mathrm{d},K}^H]^H\in\mathbb{C}^{N_{\mathrm{s}}K\times 1}$ is the transmit symbol vector with $\mathbb{E}\{\mathbf{s}_{\mathrm{d}}\mathbf{s}_{\mathrm{d}}^H\}=\mathbf{I}_{N_{\mathrm{s}}K}$, and  $\mathbf{n}_{k}\sim\mathcal{CN}(0, \sigma_{\mathrm{d}}^2\mathbf{I}_{N_{\mathrm{UE}}}) \in\mathbb{C}^{N_{\mathrm{UE}}\times 1}$ represents the noise.

Due to the constant modulus constraint, the optimization of $\boldsymbol{\phi}_{\mathrm{d}}$ is challenging. This task becomes even more complex in the system assisted by HE-IRS compared to that assisted by conventional IRS. Instead of the overall cascaded channel $\mathbf{H}_{\mathrm{ca},k}$ estimated, we rely on the DTE-based cascaded channel $\mathbf{H}_{\mathrm{ca},k}^\mathrm{DTE}$ and the STE-based equivalent channel ${\mathbf{H}_{\mathrm{eq},k}^\mathrm{STE}}$. As a result, the optimization process requires handling the interplay between the two types of channels while addressing the complexity introduced by the operation $\mathrm{mat}(\cdot)$. To tackle this, we further express the DTE-based channel $\mathrm{mat}\big(\mathbf{H}_{\mathrm{ca},k}^\mathrm{DTE} \boldsymbol{\phi}_\mathrm{d}^{*}\big)$ as
\begin{equation} \label{eqn:mat}
    \begin{array}{rl}
       \mathrm{mat}\big(\mathbf{H}_{\mathrm{ca},k}^\mathrm{DTE} \boldsymbol{\phi}_\mathrm{d}^{*}\big) = & \sum\limits_{m=1}^{N_\mathrm{DTE}}{\big[\boldsymbol{\phi}_\mathrm{d}\big]_{m}^{*}\big(\mathrm{mat}\big( \big[ \mathbf{H}_{\mathrm{ca},k}^\mathrm{DTE}\big]_{:m}\big)\big) } \\
       = &  \big[\boldsymbol{\phi}_\mathrm{d}\big]_{m_1}^{*}\big(\mathrm{mat}\big( \big[ \mathbf{H}_{\mathrm{ca},k}^\mathrm{DTE}\big]_{:m_1}\big)\big) + \sum\limits_{m_2\ne m_1}^{N_\mathrm{DTE}}\big[\boldsymbol{\phi}_\mathrm{d}\big]_{m_2}^{*}\big(\mathrm{mat}\big( \big[ \mathbf{H}_{\mathrm{ca},k}^{\mathrm{DTE}}\big]_{:m_2}\big)\big) \\
        = & \big[\boldsymbol{\phi}_\mathrm{d}\big]_{m_1}^{*} \mathbf{U}_{k,m_1}^H + \mathbf{F}_{k}^H,
    \end{array}
\end{equation}
where $\mathbf{U}_{k,m_1}^H\in\mathbb{C}^{N_\mathrm{BS} \times N_\mathrm{UE}}$ is the matrix using $\mathrm{mat}(\cdot)$ for the $m_1$-th column of $\mathbf{H}_{\mathrm{ca},k}^\mathrm{DTE}$, and $\mathbf{F}_{k}^H\in\mathbb{C}^{N_\mathrm{BS} \times N_\mathrm{UE}}$ is the sum of the product of $\big[\boldsymbol{\phi}_\mathrm{d}\big]_{m_2}^{*}$ and $\mathrm{mat}\big( \big[ \mathbf{H}_{\mathrm{ca},k}^{\mathrm{DTE}}\big]_{:m_2}\big)$ for $m_2 \ne m_1, m_2=1,\dots,N_\mathrm{DTE}$. Thus, the equivalent channel $\mathbf{H}_{\mathrm{e},k}$ can be further expressed as $\mathbf{H}_{\mathrm{e},k}=\big[\boldsymbol{\phi}_\mathrm{d}\big]_{m_1} \mathbf{U}_{k,m_1}+\mathbf{T}_k$, with $\mathbf{T}_k=\mathbf{F}_k+\big({\mathbf{H}_{\mathrm{eq},k}^\mathrm{STE}}\big)^H$. By stacking $K$ UEs as $\mathbf{U}_{m_1}=\big[\mathbf{U}_{1,m_1}^T,\dots,\mathbf{U}_{K,m_1}^T \big]^T$ and $\mathbf{T}=\big[\mathbf{T}_{1}^T,\dots,\mathbf{T}_{K}^T \big]^T$, we obtain $\mathbf{H}_\mathrm{e}=\big[\boldsymbol{\phi}_\mathrm{d}\big]_{m_1}\mathbf{U}_{m_1}+\mathbf{T}$. The effective data rate (per Hertz) of the $k$-th UE is then given by 
\begin{equation}\label{eqn:efficient SE}
\begin{array}{l}
    {R}_k=\left(1-\frac{T_\mathrm{tra}}{T_{\mathrm{tot}}}\right){{\rm{log}}\left|\mathbf{I}_{N_{\mathrm{s}}}+\mathbf{V}_{k}^H\mathbf{H}^H_{\mathrm{e},k}\boldsymbol{\Lambda}^{-1}_k\mathbf{H}_{\mathrm{e},k}\mathbf{V}_{k}\right|},
\end{array}
\end{equation}
where $\boldsymbol{\Lambda}_{k}=\sigma_{\mathrm{d}}^{2} \mathbf{I}_{N_\mathrm{UE}}+\sum_{i \neq k}^{K} {\mathbf{H}_{\mathrm{e},k}} \mathbf{V}_{i} \mathbf{V}_{i}^{H}{\mathbf{H}^H_{\mathrm{e},k}}\in\mathbb{C}^{N_{\mathrm{UE}}\times N_{\mathrm{UE}}}$ is the covariance of the noise plus interference at the $k$-th UE. Here, ${T_{\mathrm{tra}}}$ denotes the training overhead in terms of the number of pilot symbols, and ${T_{\mathrm{tot}}}$ represents the total number of symbols within the channel coherence time. 

Similar to that in \cite{2022_Lin_channel_estimation}, the SRM problem can be solved via an equivalent WMMSE problem, which can also be applied in the HE-IRS-assisted system for online beamforming optimization. Thus, the equivalent WMMSE problem is formulated as
\begin{equation}\label{eqn:WMMSE}
    \begin{array}{cl}
    \underset{{{\mathbf{V}}, \mathbf{\Phi}_{\mathrm{d}}, \mathbf{W}_k,\mathbf{\Upsilon}_k}}{\operatorname{minimize}} & {\sum_{k=1}^K\mathrm{Tr}(\mathbf{\Upsilon}_k\mathbf{E}_k)-\mathrm{log}|\mathbf{\Upsilon}_k| } \\
    \text { subject to } & {\rm{Tr}}(\mathbf{V}\mathbf{V}^H)\le{P_\mathrm{b}},\\
    &  |[\mathbf{\Phi}_{\mathrm{d}}]_{mm}| \in \mathcal{F},\ \forall m,
\end{array}
\end{equation}
where the first constraint corresponds to the transmit power constraint at the BS, with a maximum power of $P_\mathrm{b}$, and the second represents the discrete phase shift constraint of the DTEs, where $\mathcal{F}=\{ 1,{{e}^{\mathrm{j}\frac{2\pi }{2^b}}},...,{{e}^{\mathrm{j}\frac{2\pi (2^b-1)}{2^b}}} \}$ is the set of $2^b$ phase shift levels. Here, $\mathbf{E}_k=\mathbb{E}\left[\left(\mathbf{s}_{\mathrm{d},k}-\mathbf{W}^H_k\mathbf{y}_k\right)\left(\mathbf{s}_{\mathrm{d},k}-\mathbf{W}^H_k\mathbf{y}_k\right)^H\right]\in\mathbb{C}^{N_{\mathrm{s}}\times N_{\mathrm{s}}}$ is the MSE matrix of the $k$-th UE,  $\mathbf{W}_k\in\mathbb{C}^{N_{\mathrm{UE}}\times N_{\mathrm{s}}}$ is the combining matrix, and $\mathbf{\Upsilon}_k\in\mathbb{C}^{N_{\mathrm{s}}\times N_{\mathrm{s}}}$ is an auxiliary weighting matrix. In problem (\ref{eqn:WMMSE}), $\mathbf{W}_k$, $\mathbf{\Upsilon}_k$ and ${\mathbf{V}}$ can be optimized with closed-form solutions as in \cite{2024_Zhao_HEIRS}. When optimizing $\mathbf{\Phi}_{\mathrm{d}}$ while keeping $\mathbf{W}_k$, $\mathbf{\Upsilon}_k$ and ${\mathbf{V}}$ fixed, problem (\ref{eqn:WMMSE}) simplifies as follows by removing terms unrelated to $\mathbf{\Phi}_{\mathrm{d}}$.
\begin{equation}\label{eqn:WMMSE3}
    \begin{array}{cl}
    \underset{{ \mathbf{\Phi}_{\mathrm{d}}}}{\operatorname{minimize}} & \mathrm{Tr}\left(\mathbf{\Upsilon}\mathbf{W}^H\mathbf{H}_\mathrm{e}\mathbf{V}\mathbf{V}^H\mathbf{H}_\mathrm{e}^H\mathbf{W}-\mathbf{\Upsilon}\mathbf{W}^H\mathbf{H}_\mathrm{e}\right. \\
    &\left.\ \ \ \ \  \times\mathbf{V}-\mathbf{\Upsilon}\mathbf{V}^H\mathbf{H}_\mathrm{e}^H\mathbf{W}\right) \\
    \text { subject to } & |[\mathbf{\Phi}_{\mathrm{d}}]_{mm}|\in \mathcal{F},\ \forall m,
\end{array}
\end{equation}
where  $\mathbf{\Upsilon}=\mathrm{blkdiag}(\mathbf{\Upsilon}_1,\dots,\mathbf{\Upsilon}_K)$, $\mathbf{W}=\mathrm{blkdiag}(\mathbf{W}_1,\dots,\mathbf{W}_K)$, $\mathbf{E}=\mathrm{blkdiag}\big(\mathbf{E}_1,\dots,\mathbf{E}_K\b)$ and $\mathbf{H}_\mathrm{e} = \big(\mathbf{H}_{\mathrm{e},1}^T,\dots,\mathbf{H}_{\mathrm{e},K}^T\big)^T$. 

By utilizing (\ref{eqn:mat}) to extract the contribution of $\boldsymbol{\phi}_\mathrm{d}$, the EI algorithm \cite{2016_Yu_CD} can be applied to alternately optimize each element of $\mathbf{\Phi}_{\mathrm{d}}$. Therefore, the objective function of problem (\ref{eqn:WMMSE3}) can be rewritten as follows by removing terms unrelated to $\boldsymbol{\phi}_\mathrm{d}$
\begin{equation}\label{eqn:f_rewrite_Phi}
    f_\mathrm{d} = 2\mathcal{R}\{\mathrm{Tr}(\mathbf{\Upsilon}\mathbf{W}^H\mathbf{T}\mathbf{V}\mathbf{V}^H\mathbf{U}_{m_1}^H\mathbf{W}-\mathbf{\Upsilon}\mathbf{V}^H\mathbf{U}_{m_1}^H\mathbf{W})[\boldsymbol{\phi}_{\mathrm{d}}]^*_{m_1}\}.
\end{equation}
Thus, the optimal solution for $[\boldsymbol{\phi}_{\mathrm{d}}]_{m_1}$ under the constant modulus constraint is given by $[\boldsymbol{\phi}_{\mathrm{d}}]_{m_1}= -\frac{\mathrm{Tr}(\mathbf{\Upsilon}\mathbf{W}^H\mathbf{T}\mathbf{V}\mathbf{V}^H\mathbf{U}_{m_1}^H\mathbf{W}-\mathbf{\Upsilon}\mathbf{V}^H\mathbf{U}_{m_1}^H\mathbf{W})}{\big|\mathrm{Tr}(\mathbf{\Upsilon}\mathbf{W}^H\mathbf{T}\mathbf{V}\mathbf{V}^H\mathbf{U}_{m_1}^H\mathbf{W}-\mathbf{\Upsilon}\mathbf{V}^H\mathbf{U}_{m_1}^H\mathbf{W})\big|}$. As $[\boldsymbol{\phi}_{\mathrm{d}}]_{m_1}$ has finite resolution, a simple but optimal way in the sense of minimizing (\ref{eqn:f_rewrite_Phi}) is to use the close point projection method to map the resulting phase shift onto the quantized phase shift set $\mathcal{F}$. By iteratively updating the phase shifts of all DTEs, a locally optimal solution of $\boldsymbol{\phi}_{\mathrm{d}}$ is obtained. Consequently, by alternately optimizing $\mathbf{\Upsilon}$, $\mathbf{W}$, $\mathbf{V}$, and $\mathbf{\Phi}_{\mathrm{d}}$, the WMMSE-EI algorithm for online beamforming optimization is completed. 

\section{Simulation Results} \label{sec:Simulations}
\subsection{Simulation Setup}
In this section, we present extensive simulation results to evaluate the performance of the proposed channel estimation and beamforming schemes in an HE-IRS-assisted multi-user system. The BS and each UE are equipped with ULAs with $N_\mathrm{BS}=12$ and $N_\mathrm{UE}=6$, respectively. The HE-IRS is equipped with a UPA with $N=48$.
% , specifically $N_\mathrm{y}=8$ and $N_\mathrm{z}=6$ in the y- and z-directions.
Unless otherwise specified, the HE-IRS employs an equal number of DTEs and STEs, i.e., $N_\mathrm{DTE}=N_\mathrm{STE}=\frac{N}{2}$.
% , which are deployed separately on the HE-IRS, as described in Section \ref{subsec:System Model}. 
The spacing between adjacent elements in all antenna arrays and the HE-IRS is set to $\frac{\lambda}{2}$, where $\lambda=0.04\mathrm{m}$ is the wavelength of carrier wave. We assume the same number of propagation paths for both $\mathbf{G}$ and $\mathbf{H}$, i.e., $P=Q=3$, including a LoS path and two NLoS paths. A total of 3 UEs are assumed to be uniformly distributed within a specified location area, with the azimuth and elevation AoDs of the LoS path of the HE-IRS to UE channel ranging within
$\theta _\mathrm{t}^{1}\in [-45^{\circ}, 45^{\circ}]$ and $\zeta _\mathrm{t}^{1}\in [90^{\circ},180^{\circ}]$, respectively. The LoS path of the BS to HE-IRS channel is fixed, with an AoD of $\psi_\mathrm{t}^{1}=60^{\circ}$, and the azimuth and elevation AoAs of $\psi _\mathrm{r}^{1}=-60$ and $\vartheta _\mathrm{r}^{1}=120^{\circ}$. The azimuth and elevation AoAs/AoDs of the NLoS paths are randomly generated, uniformly distributed within $[-90^{\circ}, 90^{\circ}]$ and $[0,180^{\circ} ]$, respectively. The complex channel gains of the LoS paths are modeled as ${{\alpha }_{1}}({{\beta }_{1}})\sim \mathcal{CN}(0,\tau_{\mathrm{BI}}(\tau_{\mathrm{IU}}))$, while those of the NLoS paths are distributed as ${{\alpha }_{p}}({{\beta }_{q}})\sim \mathcal{CN}(0,0.1\tau_{\mathrm{BI}}(\tau_{\mathrm{IU}}))$ for $p,q\ne1$. The path loss factors $\tau_\mathrm{BI}$ and $\tau_\mathrm{IU}$ are given by $\tau_\mathrm{BI}=10^{-4.99-2\log_{10}(d_\mathrm{BI})}$, $\tau_\mathrm{IU}=10^{-4.99-2\log_{10}(d_\mathrm{IU})}$, where $d_\mathrm{BI}=25~\text{m}$ and $d_\mathrm{IU}=12~\text{m}$ denote the distance between the BS and the HE-IRS, and the HE-IRS and the UE. The phase shift precision of the DTEs is set to $b=1$ bit. $T_\mathrm{tot}$ is set to 1500, corresponding to a channel coherence time of 5ms with a transmission bandwidth of $3\times10^5$ Hz.  The uplink training pilot-to-noise-ratio (PNR) is defined as $\frac{P_{\mathrm{tr}}\tau_{\mathrm{BI}}\tau_{\mathrm{IU}}}{\sigma^2}$, while the downlink transmission signal-to-noise-ratio (SNR) is defined as $\frac{P_\mathrm{b}\tau_{\mathrm{BI}}\tau_{\mathrm{IU}}}{\sigma_\mathrm{d}^2}$.

\subsection{Performance of the DSD-MO Channel Estimation Scheme}
As discussed in Section \ref{subsec:The Channel Estimation Challenge for HE-IRS}, the overall cascaded channel in HE-IRS-assisted systems is not suitable for evaluating channel estimation performance. Instead, we utilize the two evaluation channels: The DTE-based cascaded channel $\mathbf{H}_\mathrm{ca,k}^\mathrm{DTE}=\mathbf{H}_\mathrm{DTE,k}^T \otimes \mathbf{G}_\mathrm{DTE}$, and the STE-based equivalent channel $\mathbf{H}_\mathrm{eq,k}^\mathrm{STE}=\mathbf{G}_\mathrm{STE} \boldsymbol{\Omega} \mathbf{H}_\mathrm{STE,k}$. The average normalized MSE (NMSE) of these channels across all UEs serves as the performance metric for HE-IRS channel estimation, which are defined as
% As discussed in Section \ref{subsec:The Channel Estimation Challenge for HE-IRS}, the overall cascaded channel in HE-IRS-assisted systems is found to be unsuitable for evaluating channel estimation performance using $\mathbf{H}^T \otimes \mathbf{G}$. Instead, as discussed in Section \ref{subsec:The DTE-STE Decoupled Channel and its Sparsity}, we should utilize the DTE-based cascaded channel $\mathbf{H}_\mathrm{ca,k}^\mathrm{DTE}=\mathbf{H}_\mathrm{DTE,k}^T \otimes \mathbf{G}_\mathrm{DTE}$ and the STE-based equivalent channel $\mathbf{H}_\mathrm{eq,k}^\mathrm{STE}=\mathbf{G}_\mathrm{STE} \boldsymbol{\Omega} \mathbf{H}_\mathrm{STE,k}$. Consequently, we take the average normalized MSE (NMSE) of both $\mathbf{H}_\mathrm{ca,k}^\mathrm{DTE}$ and $\mathbf{H}_\mathrm{eq,k}^\mathrm{STE}$ across all UEs as the performance metric for evaluating the channel estimation performance of the HE-IRS-assisted system, which are defined as
% \begin{equation} \label{eqn:NMSE of DTE}
%     \mathrm{NMSE}_\mathrm{ca,k}^\mathrm{DTE}=\mathbb{E}\left\{\frac{\sum_{k=1}^K\big\|\big(\mathbf{H}_\mathrm{DTE,k}^T \otimes \mathbf{G}_\mathrm{DTE}\big)-\big(\widehat{\mathbf{H}}_\mathrm{DTE,k}^T\otimes\widehat{\mathbf{G}}_\mathrm{DTE}\big)\big\|^2_{F}}{\sum_{k=1}^K\big\|\mathbf{H}_\mathrm{DTE,k}^T \otimes \mathbf{G}_\mathrm{DTE}\big\|^2_F} \right\},
% \end{equation}
\begin{equation} \label{eqn:NMSE of DTE}
    \mathrm{NMSE}_\mathrm{DTE,ca}=\mathbb{E}\left\{\frac{1}{K}\sum_{k=1}^K{\frac{\big\|\mathbf{H}_\mathrm{ca,k}^\mathrm{DTE}-\widehat{\mathbf{H}}_\mathrm{ca,k}^\mathrm{DTE}\big\|^2}{\big\|\mathbf{H}_\mathrm{ca,k}^\mathrm{DTE}\big\|^2}} \right\},
\end{equation}
\begin{equation} \label{eqn:NMSE of STE}
    \mathrm{NMSE}_\mathrm{STE,eq}=\mathbb{E}\left\{\frac{1}{K}\sum_{k=1}^K\frac{\big\|\mathbf{H}_\mathrm{eq,k}^\mathrm{STE}-\widehat{\mathbf{H}}_\mathrm{eq,k}^\mathrm{STE}\big\|^2}{\big\|\mathbf{H}_\mathrm{eq,k}^\mathrm{STE}\big\|^2} \right\}.
\end{equation}

To provide a performance benchmark, we modify  a three-stage channel estimation scheme in conventional IRS-assisted systems \cite{2022_Lin_channel_estimation} and apply it to HE-IRS-assisted systems. This scheme sequentially estimates the AoDs at the UE, the AoAs at the BS, and the cascaded channel in each stage using an OMP algorithm. The overall cascaded channel is then decomposed into the DTE-based cascaded channel and the STE-based equivalent channel. In particular, the angular resolutions at the BS, UE, HE-IRS are set to $G_\mathrm{BS}=128$, $G_\mathrm{UE}=128$, $G_\mathrm{I}=1024$, respectively.
% Thus, the performance metric for HE-IRS channel estimation in \ref{eqn:NMSE of DTE} and \ref{eqn:NMSE of STE} can be utilized. 
This modified scheme is referred to as CS-EST. 

\begin{figure}
	\centering
    \captionsetup{font={small}}
    \includegraphics[height=7cm,width=10cm]{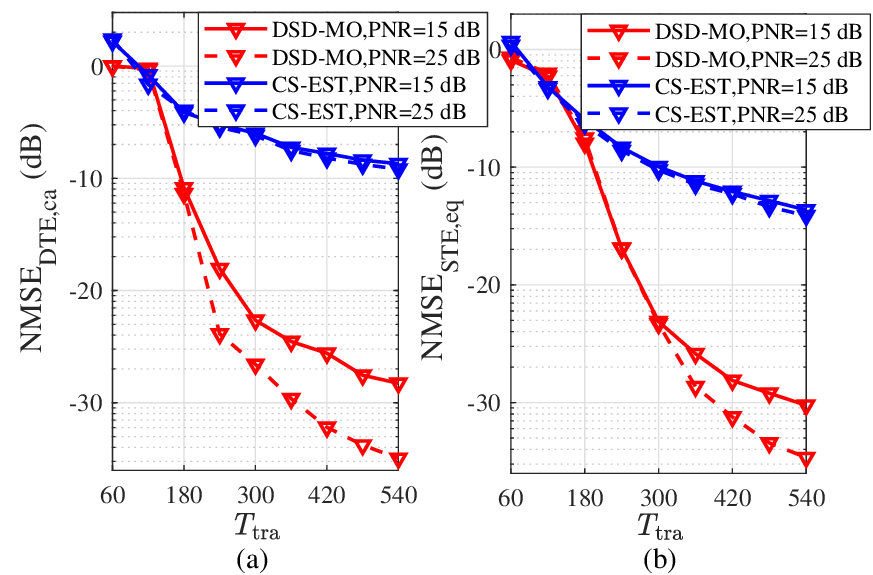}
	\caption{(a) $\mathrm{NMSE}_\mathrm{DTE,ca}$ vs. $T_\mathrm{tra}$ for different schemes and PNRs in the estimation of the DTE-based cascaded channel. (b) $\mathrm{NMSE}_\mathrm{STE,eq}$ vs. $T_\mathrm{tra}$ for different schemes and PNRs in the estimation of  the STE-based equivalent channel.}	\label{fig:Fig1_NMSE_T}
\end{figure}

In Fig. \ref{fig:Fig1_NMSE_T}, we first evaluate the NMSE performance as a function of the total pilot overhead of all UEs, $T_\mathrm{tra}$, for different estimation schemes and PNRs, with $N=48$. As shown in Fig. \ref{fig:Fig1_NMSE_T}(a) and \ref{fig:Fig1_NMSE_T}(b), for $\mathrm{NMSE}_\mathrm{DTE,ca}$ in the DTE-based cascaded channel and $\mathrm{NMSE}_\mathrm{STE,eq}$ in the STE-based equivalent channel, the performance of both DSD-MO and CS-EST schemes improves as $T_\mathrm{tra}$ increases. For the CS-EST scheme, the performance saturates as $T_\mathrm{tra}$ and PNR increase, due to its estimation solution being constrained within a pre-defined feasible set. In contrast, the proposed DSD-MO scheme initially suffers from insufficient training overhead caused by the limited number of observations. However, as $T_\mathrm{tra}$ increases, its performance improves rapidly before reaching the slight descent region. Additionally, the performance of the DSD-MO scheme continues to improve with higher PNR.
\begin{figure}
	\centering
    \captionsetup{font=small}
    \includegraphics[height=7cm,width=10cm]{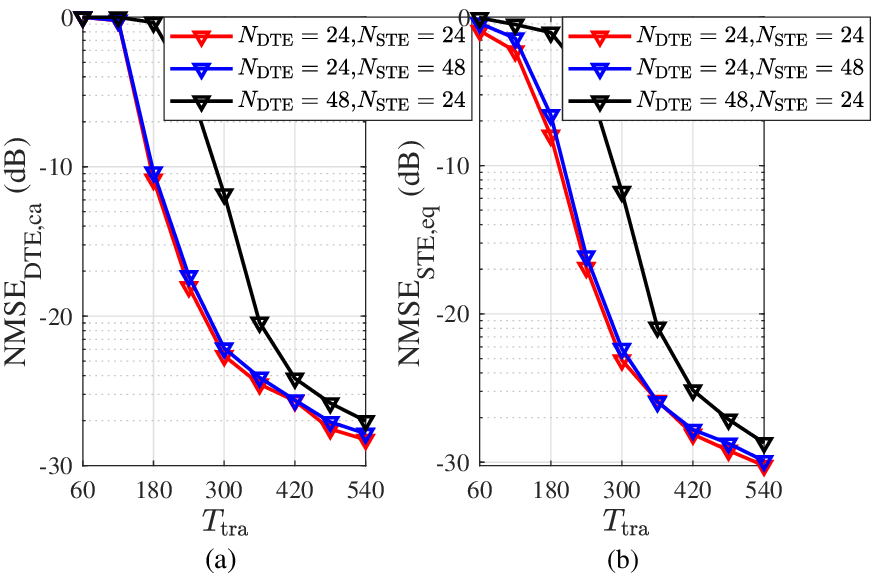}
	\caption{(a) $\mathrm{NMSE}_\mathrm{DTE,ca}$ vs. $T_\mathrm{tra}$ for different HE-IRS configurations in the DTE-based cascaded channel. (b) $\mathrm{NMSE}_\mathrm{STE,eq}$ vs. $T_\mathrm{tra}$ for different configurations in the STE-based equivalent channel.}	\label{fig:Fig2_NMSE_T}
\end{figure}

In Fig. \ref{fig:Fig2_NMSE_T}, we further demonstrate the NMSE performance for the DSD-MO scheme with different HE-IRS configurations and $T_\mathrm{tra}$, with $\mathrm{PNR}=15$ dB. There are three HE-IRS configurations considered: (1) $N_\mathrm{DTE}=24, N_\mathrm{STE}=24$, (2) $N_\mathrm{DTE}=24, N_\mathrm{STE}=48$, and (3) $N_\mathrm{DTE}=48, N_\mathrm{STE}=24$. It can be observed that, the performance of the second configuration with double STEs numbers is very close to that of the first configuration across all $T_\mathrm{tra}$. However, the third configuration with double DTEs numbers requires more pilot overhead to achieve significant performance improvements and reaches the followed slight descent region. A crucial reason is that the number of variables to be estimated in DTE-based cascaded channel ($\mathbf{H}_\mathrm{ca,k}^\mathrm{DTE}\in\mathbb{C}^{N_\mathrm{BS}N_\mathrm{UE} \times N_\mathrm{DTE}}$) is proportional to the number of DTEs, while the STE-based equivalent channel ($\mathbf{H}_\mathrm{eq,k}^\mathrm{STE}\in\mathbb{C}^{N_\mathrm{BS} \times N_\mathrm{UE}}$) is not directly related to the number of STEs.
\begin{figure}
	\centering
    \captionsetup{font=small}
    \includegraphics[height=6.7cm,width=7.9cm]{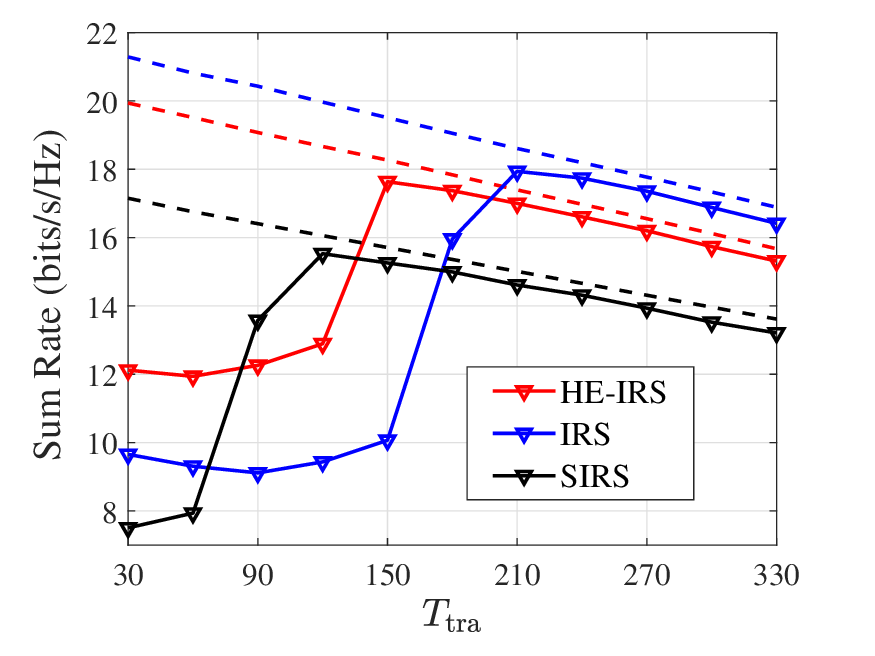}
	\caption{Sum rate performance vs. $T_\mathrm{tra}$ for HE-IRS-, IRS-, SIRS-assisted systems with perfect CSI (dashed lines) and estimated CSI (solid lines).}	\label{fig:Fig3_Rate_T}
\end{figure}

\subsection{Comparison of the Sum Rate Performance for HE-IRS-, IRS-, SIRS-Assisted Systems}
In this subsection, we compare the sum rate performance of the HE-IRS-assisted system to that of the conventional IRS- and SIRS-assisted systems with both perfect CSI and estimated CSI. 
% When the number of STEs decreases to zero, the online channel estimation and beamforming methods for HE-IRS, namely DSD-MO and WMMSE-EI, degrade to channel estimation and beamforming methods for systems assisted by the surface containing only DTEs. 
To ensure a fair comparison, the channel estimation and beamforming algorithms for conventional IRS- and SIRS-assisted systems are adapted from the HE-IRS algorithms when the number of STEs is reduced to zero. Additionally, for the conventional IRS, all of $N$ reflecting elements are DTEs. For the SIRS, it is assumed to have $\frac{N}{2}$ DTEs on its surface. All other simulation parameters remain identical.

In Fig. \ref{fig:Fig3_Rate_T}, we show the sum rate performance as a function
of $T_\mathrm{tra}$ for HE-IRS-, IRS-, SIRS-assisted systems with perfect CSI (dashed lines) and estimated CSI (solid lines), when $N=48$, $\mathrm{PNR}=15$ dB, and $P_\mathrm{b}=30$ dBm. For all systems, the sum rate performance with perfect CSI decreases monotonically as $T_\mathrm{tra}$ increases, due to the reduced time available for data transmission according to (\ref{eqn:efficient SE}). Meanwhile, the sum rate performance with estimated CSI all initially suffers a severe degradation at small $T_\mathrm{tra}$ due to the low channel estimation quality, and starts to increase rapidly when $T_\mathrm{tra}$ becomes larger. As $T_\mathrm{tra}$ continues to increase, the performance begins to decrease as the estimation quality cannot be significantly improved. Each system exhibits a best choice of $T_\mathrm{tra}$ to balance the sum rate improvement from accurate channel estimation and the data rate loss from pilot overhead. The best $T_\mathrm{tra}$ differs across systems, e.g., 120 for SIRS, 150 for HE-IRS, and 210 for IRS. These differences are primarily due to the varying number of variables to be estimated. For SIRS, only the cascaded channel for half DTEs needs to be estimated. For HE-IRS, the cascaded channel for half DTEs and the equivalent channel for half STEs need to be estimated. For IRS, the cascaded channel for all DTEs needs to be estimated, which results in the largest number of variables.
\begin{figure}
	\centering
    \captionsetup{font=small}
    \includegraphics[height=6.7cm,width=7.9cm]{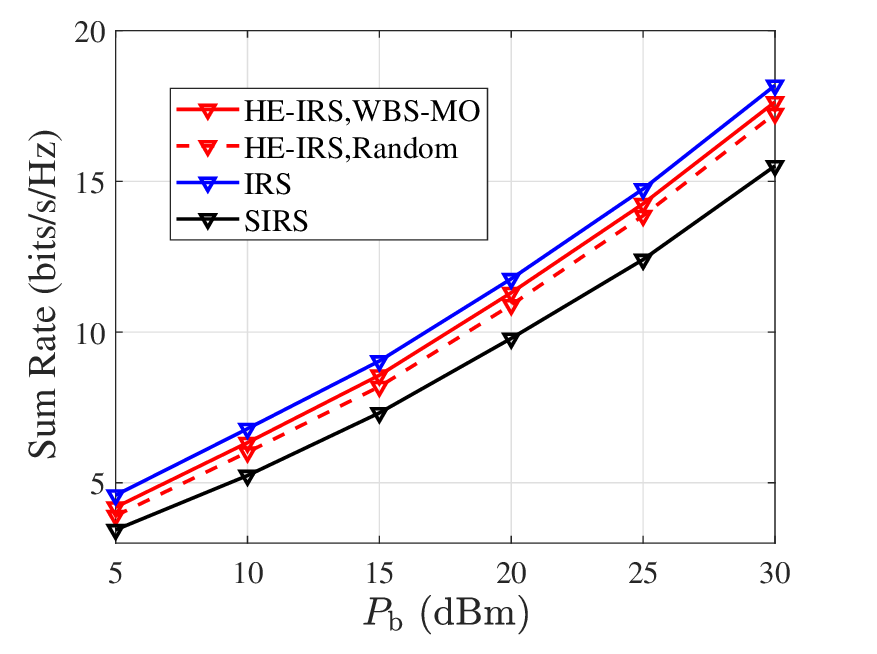}
	\caption{Sum rate performance vs. $P_\mathrm{b}$ for HE-IRS-, IRS-, SIRS-assisted systems with estimated CSI.}	\label{fig:Fig4_Rate_Pb}
\end{figure}

In Fig. \ref{fig:Fig4_Rate_Pb}, we evaluate the sum rate performance as a function
of $P_\mathrm{b}$ for HE-IRS-, IRS-, SIRS-assisted systems with estimated CSI. Here, $N=48$, $\mathrm{PNR}=\mathrm{SNR}+15$ dB, and the best $T_\mathrm{tra}$ for each system and each PNR is selected to maximize the sum rate. As observed, the sum rate performance of all systems increases with higher $P_\mathrm{b}$. Specifically, compared with random phase shifts of the STEs, the sum rate performance of the HE-IRS improves by optimizing their phase shifts to concentrate power on the UE location area in the offline stage. The conventional IRS always achieves the highest sum rate performance, as all of its elements are DTEs. Although the SIRS requires less pilot overhead for channel estimation, it suffers significant performance losses with only half DTEs. For the proposed HE-IRS with half DTEs and STEs, its performance is much closer to the conventional IRS for all $P_\mathrm{b}$. This excellent performance not only results from the two-stage beamforming design, but also results from the reduction of required pilot overhead by efficiently exploiting the unique HE-IRS structure characteristics for channel estimation.
% In Fig. \ref{fig:Fig_Stream}, we show $\mathrm{NMSE}_\mathrm{STE,eq}$ versus actual path number for the STE-based equivalent channel.
% \begin{figure*}[!t]
%     \centering
%     \captionsetup{font=small}
%     \subfigure[Caption for Fig. A]{
%         \includegraphics[width=0.3\textwidth]{Fig_Stream_SIRS.eps}
%         \label{fig:a}
%     }\hfill
%     \subfigure[Caption for Fig. B]{
%         \includegraphics[width=0.3\textwidth]{Fig_Stream_HEIRS.eps}
%         \label{fig:b}
%     }\hfill
%     \subfigure[Caption for Fig. C]{
%         \includegraphics[width=0.3\textwidth]{Fig_Stream_IRS.eps}
%         \label{fig:c}
%     }
%     \caption{(a) The reflected beam patterns of the STEs. (b) The reflected beam patterns of the DTEs. (c) The reflected beam patterns of the HE-IRS.}
%     \label{fig:Fig_Stream}
% \end{figure*}
\begin{figure}
	\centering
    \captionsetup{font=small}
    \includegraphics[height=6.7cm,width=7.9cm]{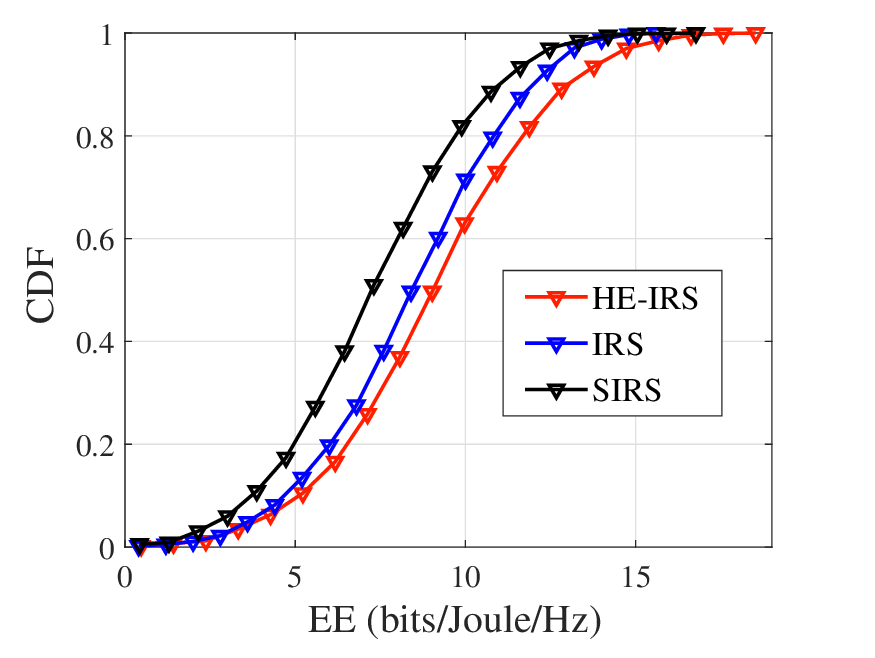}
	\caption{The CDF of the EE performance for HE-IRS-, IRS-, SIRS-assisted systems with estimated CSI.}	\label{fig:Fig5_CDF_EE}
\end{figure}
\subsection{Comparison of the Energy Efficiency Performance of HE-IRS-, IRS-, SIRS-Assisted Systems}
In this section, we evaluate the EE performance for the three systems with estimated CSI. Following the definition in \cite{2022_WCL_Zargari_EE_use}, the EE is calculated as $\mathbb{E}\big\{\sum_{k=1}^K R_k/{{P}_\mathrm{total}}\big\}$, where ${{P}_\mathrm{total}}$ $={{P}_\mathrm{b}}$ $+{{P}_\mathrm{cir}}$ $+{{P}_\mathrm{IRS}}$ is the total consumed power of the system.  ${{P}_\mathrm{cir}}$ $=K P_\mathrm{UE}^\mathrm{cir}$ $+P_\mathrm{BS}^\mathrm{cir}$ denotes the circuit power of the BS ($P_\mathrm{BS}^\mathrm{cir}$ = 30 dBm) and the UE ($P_\mathrm{UE}^\mathrm{cir}$ = 15 dBm). The power consumption of the PIN-based IRS is given by $P_{\mathrm{IRS}}$ $=P_\mathrm{static}$ $+\sum\nolimits_{{m}=1}^{{{N_\mathrm{DTE}}}}{{{t}_{{{m}}}} \cdot P_{\mathrm{PIN}}}$, where $P_\mathrm{static}=15$ dBm is the static power consumption, $t_{m}$ is the number of the on-state PIN diodes for the $m$-th DTE, and $P_{\mathrm{PIN}}=12$ dBm is the power consumption of each on-state PIN diode \cite{2024_Jin_Power}. In Fig. \ref{fig:Fig5_CDF_EE}, we show the cumulative distribution function (CDF) of the EE over random UEs locations, with $N=48$, $\mathrm{PNR}=15$ dB, and $P_\mathrm{b}=25$ dBm. The best $T_\mathrm{tra}$ for each system is selected. As observed, unlike the sum rate performance in Fig. \ref{fig:Fig4_Rate_Pb}, the HE-IRS outperforms both the conventional IRS and the SIRS in terms of EE, attributed to integrating half of its elements with power-free STEs.
\subsection{Performance of the DSD-MO Channel Estimation Scheme with Channel Rank Mismatches}
As described in Section \ref{subsec:Rank Selection Rule with Mismatched Path Numbers}, when the number of the propagation paths is not perfectly known, the mismatched path number results in mismatches between the estimated channel rank and the actual channel rank, which necessitates a robust rank selection rule developed for the MO method. For comparison, we take the DSD-MO scheme that directly uses the estimated rank, referred to as E-rank, while the DSD-MO scheme that uses the selected rank is referred to as S-rank. 

\begin{figure}
	\centering
    \captionsetup{font=small}
    \includegraphics[height=7cm,width=10cm]{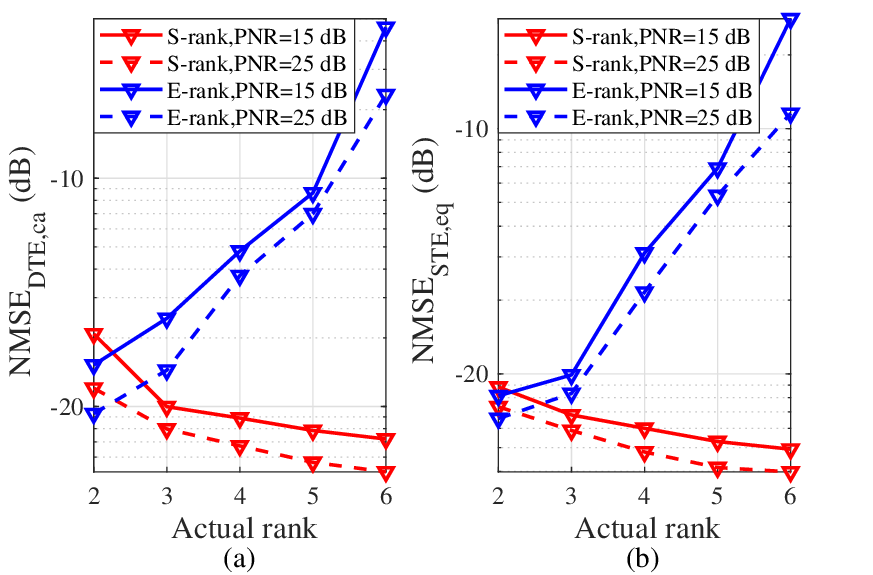}
	\caption{(a) $\mathrm{NMSE}_\mathrm{DTE,ca}$ vs. actual rank of the MO method for the DTE-based cascaded channel. (b) $\mathrm{NMSE}_\mathrm{STE,eq}$ vs. actual rank of the MO method for the STE-based equivalent channel.}	\label{fig:Fig6_NMSE_T}
\end{figure}

In Fig. \ref{fig:Fig6_NMSE_T}, we demonstrate the NMSE performance as a function of the actual rank for the proposed DSD-MO channel estimation scheme, with the estimated rank varying from 2 to 6. We set $N=48$, $\mathrm{PNR}=15$ dB, and $T_\mathrm{tra}=360$. The maximum actual rank, i.e., 6, is  chosen as the selected rank. As observed, for both $\mathrm{NMSE}_\mathrm{DTE,ca}$ and $\mathrm{NMSE}_\mathrm{STE,eq}$, the performance of E-rank deteriorates significantly as the actual rank increases. However, the performance of S-rank improves gradually with increasing actual rank. This aligns with Lemmas 2 and 3, which indicate that using a rank lower than the true rank as a fixed-rank constraint restricts the ability to approximate the actual channels. On the other hand, selecting a rank higher than the true rank retains the ability to approximate the actual channels arbitrarily closely,  thereby ensuring robust performance under channel rank mismatches.
\section{Conclusion} \label{sec:Conclusion}
In this paper, we have investigated the channel estimation and beamforming problems in the HE-IRS-assisted multi-user MIMO system. We showed that the overall cascaded channel estimation cannot be directly applied to the unique DTE-STE integrated structure, and then proposed the DSD-MO channel estimation scheme, which efficiently leverages the characteristics of the HE-IRS structure, and the inherent sparsity in both DTE-based cascaded and STE-based equivalent channels. To further mitigate channel rank mismatches caused by imperfect sparsity information, we developed a robust rank selection rule. As for beamforming optimization, we proposed the offline WBS-MO beamforming algorithm for the STEs to achieve wide beam coverage across the UE location area. Based on the HE-IRS channel estimated by the DSD-MO scheme and the STE phase shifts optimized by the WBS-MO algorithm, we proposed the online WMMSE-EI algorithm for the beamforming of the BS and the DTEs. Simulation results demonstrated the effectiveness of the proposed channel estimation and beamforming schemes. Compared with the conventional IRS with the same size, the HE-IRS not only reduces power consumption but also requires less pilot overhead. With the proposed schemes, the green HE-IRS achieves competitive sum rate performance as that of the conventional IRS but with higher EE performance. 

\end{document}